\documentclass[sort&compress,final,3p,times,twocolumn,fixfloat]{elsarticle}

\usepackage{graphicx,color}

\journal{Physics Letters B}

\newcommand{\eqs}[1]{\begin{equation} \begin{split} #1\end{split} \end{equation} }

\def\ie{{\it i.e.}}
\def\eg{{\it e.g.}}
\def\etal{{\it et al.}}

\newcommand{\ce}[1]{Eq.~(\ref{#1})}
\newcommand{\cf}[1]{{Fig.~\ref{#1}}}

\usepackage{txfonts}
\usepackage{bm}
\usepackage{amsmath}
\usepackage{amssymb}
\usepackage{booktabs}
\usepackage{float}

\usepackage{xspace} 
\usepackage{mathtools, cuted}

\usepackage{scalerel}
\usepackage{tikz}
\usetikzlibrary{svg.path}

\definecolor{orcidlogocol}{HTML}{A6CE39}
\tikzset{
  orcidlogo/.pic={
    \fill[orcidlogocol] svg{M256,128c0,70.7-57.3,128-128,128C57.3,256,0,198.7,0,128C0,57.3,57.3,0,128,0C198.7,0,256,57.3,256,128z};
    \fill[white] svg{M86.3,186.2H70.9V79.1h15.4v48.4V186.2z}
                 svg{M108.9,79.1h41.6c39.6,0,57,28.3,57,53.6c0,27.5-21.5,53.6-56.8,53.6h-41.8V79.1z M124.3,172.4h24.5c34.9,0,42.9-26.5,42.9-39.7c0-21.5-13.7-39.7-43.7-39.7h-23.7V172.4z}
                 svg{M88.7,56.8c0,5.5-4.5,10.1-10.1,10.1c-5.6,0-10.1-4.6-10.1-10.1c0-5.6,4.5-10.1,10.1-10.1C84.2,46.7,88.7,51.3,88.7,56.8z};
  }
}

\newcommand\orcidicon[1]{\href{https://orcid.org/#1}{\mbox{\scalerel*{
\begin{tikzpicture}[yscale=-1,transform shape]
\pic{orcidlogo};
\end{tikzpicture}
}{|}}}}

\newcommand{\GeV}{\,\text{GeV}}

\newcommand{\Q}{\mathcal{Q}}

\usepackage[normalem]{ulem}

\newcommand{\NLOstar}{NLO$^\star\,$}
\def\jpsi    {\mbox{$J/\psi$}}

\newcommand{\lf}{\left}
\newcommand{\rg}{\right}

\newcommand*\oline[1]{%
   \vbox{%
     \hrule height 0.5pt
     \kern0.4ex
     \hbox{%
       \kern-0.15em
       \ifmmode#1\else\ensuremath{#1}\fi
       \kern-0.15em
     }
   }
}

 \usepackage{ifpdf}
 \ifpdf
 \usepackage[pdftex]{hyperref}
 \else
 \usepackage[hypertex]{hyperref}
 \fi

 \hypersetup{
   pdftitle={},%
   pdfauthor={},%
   pdfsubject={},%
   pdfkeywords={},%
   pdfstartview={},%
   bookmarksopen=true, breaklinks=true, debug=true, %
   colorlinks=true, linkcolor=red, citecolor=blue, urlcolor=blue
 }

\usepackage{float} 
\usepackage[caption=false]{subfig} 

\usepackage[sort&compress, capitalise]{cleveref}

\newcount\savefnused
\newcount\savefndone

\newcommand{\savefootnote}[2][\empty]
{\ifx\empty#1\footnotemark\else\footnotemark[#1]\fi
 \global\advance\savefnused by 1
 \expandafter\xdef\csname savefnmark\the\savefnused\endcsname{\thefootnote}%
 \expandafter\xdef\csname savefntext\the\savefnused\endcsname{#2}%
}
\newcommand{\flushfootnote}{\loop\ifnum\savefndone<\savefnused
  \global\advance\savefndone by 1
  \footnotetext[\csname savefnmark\the\savefndone\endcsname]%
    {\csname savefntext\the\savefndone\endcsname}%
  \global\expandafter\let\csname savefnmark\the\savefndone\endcsname\relax
  \global\expandafter\let\csname savefntext\the\savefndone\endcsname\relax
\repeat}

\begin{document}

\title{Large-$P_T$ inclusive photoproduction of $J/\psi$  in electron-proton collisions at HERA and the EIC}

\date{\today}

\address[add1]{Universit\'e Paris-Saclay, CNRS, IJCLab, 91405 Orsay, France}
\address[add2]{Laboratoire de Physique Th\'eorique et Hautes Energies, UMR 7589,
Sorbonne Universit\'e et CNRS, 4 place Jussieu, 75252 Paris Cedex 05, France}
\address[add3]{V.N. Karazin Kharkiv National University Research institute the School of Physics and Technology, 4 Svobody Sq., 61022, Kharkiv, Ukraine}

\author[add1]{Carlo Flore\corref{cor1}\orcidicon{0000-0002-1071-204X}}
\ead{carlo.flore@ijclab.in2p3.fr}
\cortext[cor1]{Corresponding author}

\author[add1]{Jean-Philippe Lansberg\orcidicon{0000-0003-2746-5986}}
\ead{jean-philippe.lansberg@in2p3.fr}

\author[add2]{Hua-Sheng Shao\orcidicon{0000-0002-4158-0668}}
\ead{huasheng.shao@lpthe.jussieu.fr}

\author[add1,add3]{Yelyzaveta Yedelkina\orcidicon{0000-0002-2366-9001}}
\ead{yelyzaveta.yedelkina@universite-paris-saclay.fr}

\begin{abstract}
We study the inclusive $J/\psi$ production at large transverse momenta at lepton-hadron colliders in the limit when the exchange photon is quasi real, also referred to as photoproduction. Our computation includes the leading-$P_T$ leading-$v$ next-to-leading $\alpha_s$ corrections. In particular, we consider the contribution from $J/\psi$ plus another charm quark, by employing for the first time in quarkonium photoproduction the variable-flavour-number scheme. We also include a QED-induced contribution via an off-shell photon which remained ignored in the literature and which we show to be the leading contribution at high $P_T$ within the reach of the EIC. In turn, we use our computation of $J/\psi+$charm to demonstrate its observability at the future EIC and the EIC sensitivity to probe the non-perturbative charm content of the proton at high $x$.
\end{abstract}

\maketitle

\section{\label{sec:intro}Introduction}
\vspace*{-0.25cm}
Quarkonium inclusive reactions have been widely studied at colliders since the discoveries of the first quarkonia in the 70's but they remain unsatisfactorily understood. We guide the reader to the following reviews covering the HERA and Tevatron legacy~\cite{Kramer:2001hh,Brambilla:2004wf,Lansberg:2006dh,Brambilla:2010cs} and the current evolution of the field at RHIC and the LHC~\cite{Andronic:2015wma,Lansberg:2019adr}.

Inclusive $\jpsi$ photoproduction, when a near on-shell photon hits and breaks a proton to produce the $\jpsi$, has been the object of several studies at HERA~\cite{Aid:1996dn,Breitweg:1997we,Chekanov:2002at,Adloff:2002ex,Chekanov:2009ad,Aaron:2010gz,Abramowicz:2012dh} in order to understand the quarkonium-production mechanisms and with the hope to learn more about the gluon content of the proton (see \eg~\cite{Jung:1992uj}). Indeed, the replacement of one hadron by a photon, in principle, simplifies the computations of such direct photon-proton interactions with respect to hadron-hadron ones. This also allows one to safely use quarkonia to extract Transverse Momentum Dependent gluon distributions~\cite{Bacchetta:2018ivt,DAlesio:2019qpk,Kishore:2019fzb,Boer:2020bbd}.

Otherwise, this process is very similar to the widely studied hadroproduction where two hadrons collide to produce the quarkonium. Yet, the photoproduction cross sections are smaller and require significant luminosities to produce objects like quarkonia. The statistical samples collected at HERA are indeed more limited with quasi no data on $\psi'$ and none on the bottomonia.

Besides, at high energies, the hadronic content of the photon can be ``resolved'' by these reactions. The computations relevant for these interactions between a resolved-photon and a proton are then similar to those for hadroproduction. Moreover, the inclusive contribution is not always much larger than the exclusive or diffractive ones. Indeed, for vector mesons like the $J/\psi$, the corresponding cross sections can be extremely large. Both resolved-photon and diffractive contributions can be avoided with specific kinematical cuts as we shall see later. A last possible complication comes from the possible decays --also referred to as feed down (FD)-- of excited quarkonia or $b$ quarks, which should be analysed case by case like for hadroproduction.

In the present analysis, we focus on the large-$P_T$ region ($P_T \gg M_{J/\psi }$) which was seldom studied at HERA and which could be experimentally studied in more details at the future US Electron-Ion Collider (EIC)~\cite{Accardi:2012qut}. We will use the Colour-Singlet Model (CSM)~\cite{Berger:1980ni}, which is the leading-$v$ contribution of Non-Relativistic QCD (NRQCD)~\cite{Bodwin:1994jh}. Since the seminal work by Kr\"amer~\cite{Kramer:1995nb} when he found that real-emission $\alpha\alpha_s^3$ corrections associated to $t$-channel-gluon-exchange topologies were large, we know that such NLO CSM corrections are in general large when $P_T$ increases. This can be traced back to the more favourable $P_T$ scaling of these specific real-emission contributions. The same has been observed in several hadroproduction processes of spin-triplet vector quarkonia~\cite{Artoisenet:2008fc,Lansberg:2008gk,Lansberg:2009db,Gong:2012ah,Lansberg:2013qka,Lansberg:2014swa}. The conclusion made by Kr\"amer that
the HERA data could be accounted for by the CSM after the inclusion of the NLO corrections were then questioned by
3 studies~\cite{Artoisenet:2009xh,Butenschoen:2009zy,Chang:2009uj} in particular at large $P_T$. We will revisit this for HERA in  order to provide predictions for the EIC.

If the analogy holds with hadroproduction, the leading-$P_T$ fragmentation processes, scaling like $P_T^{-4}$, should provide the bulk
of the large-$P_T$ yield. One notable difference is that fragmentation cannot be initiated by $g+g \to g+g$ partonic reactions. 
Quarks are necessarily involved, which reduces their overall impact at HERA energies. In 1996, Godbole \etal\ \cite{Godbole:1995ie} further showed that the charm fragmentation contribution was taking over the gluon fragmentation for $P_T> 10$ GeV (see also~\cite{Kniehl:1997gh}) which leads us to reconsider the contribution of $J/\psi+\text{(an unobserved) }c$ to the inclusive production. Another $J/\psi+c$ channel, where a $c$-quark comes from the proton, had already been addressed in the early 80's by Berger and Jones~\cite{Berger:1982fh} but was essentially forgotten. We also revisit it to show that it can be a relevant contribution to the inclusive yield but also that it can be studied for itself by further detecting a charmed hadron in the final state.

We also consider another overlooked contribution, namely from the pure QED process where the $J/\psi$ is produced by an off-shell photon, which happens to be relevant for the large-$P_T$ domain. Owing to the presence of a quark line interacting with the initial photon, such a contribution should be as large as the typical fragmentation Colour-Octet (CO) contributions~\cite{Lansberg:2013wva}. As a matter of fact, we see that this QED contribution matters when one reaches the valence region in the proton. 

Thanks to different possible running energies, the EIC will indeed allow one to probe rather large $x$. This also motivate us to see to which extent extracting the $J/\psi+c$ yield in this region could help measuring a valence-like non-perturbative charm content in the proton, also referred to as Intrinsic Charm (IC)~\cite{Brodsky:1980pb}. 

The structure of our Letter is as follows. Section~\ref{sec:methods} outlines our methodology to compute the contributions of the aforementioned reactions, whereas Section~\ref{sec:digression-FD} focusses on the FD from excited quarkonium states or $b$ decays. Section~\ref{sec:results} gathers our results for the inclusive cross section for the HERA and EIC kinematics, while
Section~\ref{subsec:charm} is devoted to the discussion of $J/\psi+$charm associated production. Section~\ref{sec:conclusions} gathers our conclusions.

\vspace*{-0.35cm}
\section{\label{sec:methods}$J/\psi$ photoproduction at finite $P_T$}
\vspace*{-0.25cm}

As announced, we will limit our study to the CSM.  
As such, at Born order, or LO in $\alpha_s$, $J/\psi$ photoproduction proceeds via the partonic process
$\gamma+g \to J/\psi+g$ thus at $\alpha \alpha_s^2$ (\cf{gammag-psig-LO-CSM}). Strictly speaking, $\gamma+ \{c,\bar c\} \to J/\psi+\{c,\bar c\}$ (\cf{gammac-psic-CSM})  appears at the same order in $\alpha_s$ as far as the matrix element is concerned~\cite{Berger:1982fh}.

\begin{figure}[hbt!]
\centering
\subfloat[]{\includegraphics[scale=.25]{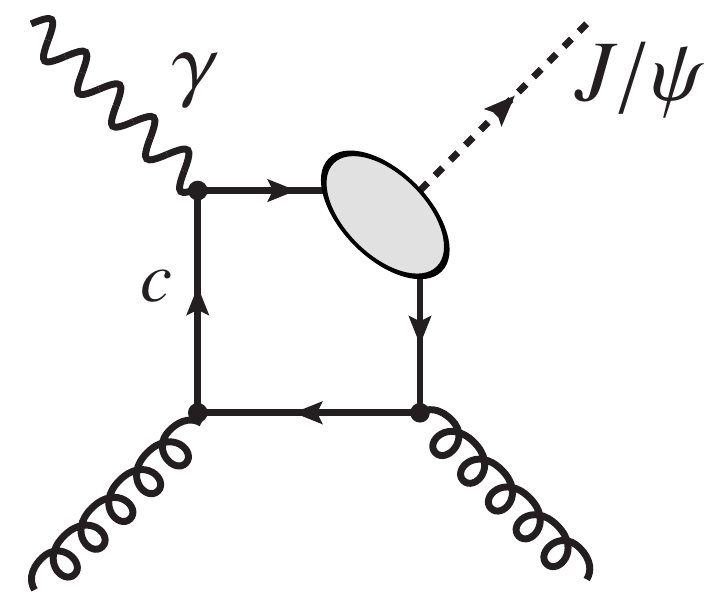}\label{gammag-psig-LO-CSM}}
\subfloat[]{\includegraphics[scale=.25]{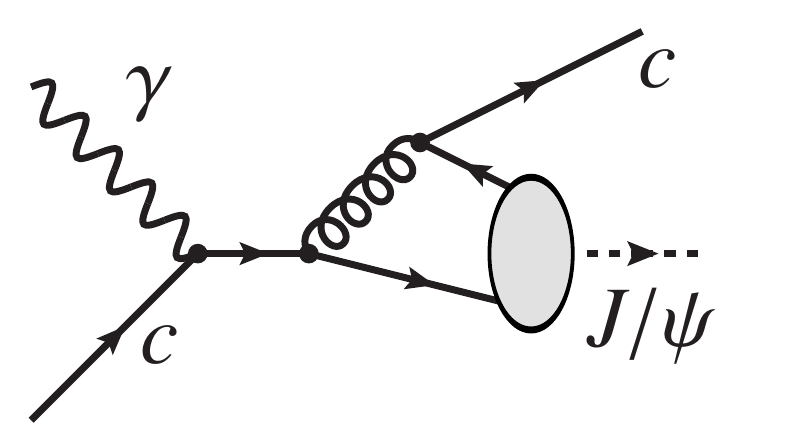}\label{gammac-psic-CSM}}
\subfloat[]{\includegraphics[scale=.25]{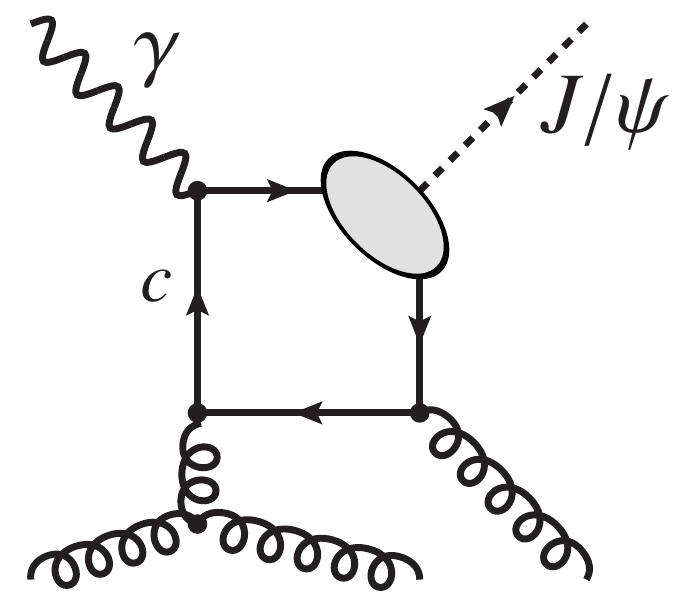}\label{gammag-psig-NLO-PT6-CSM}}
\subfloat[]{\includegraphics[scale=.25]{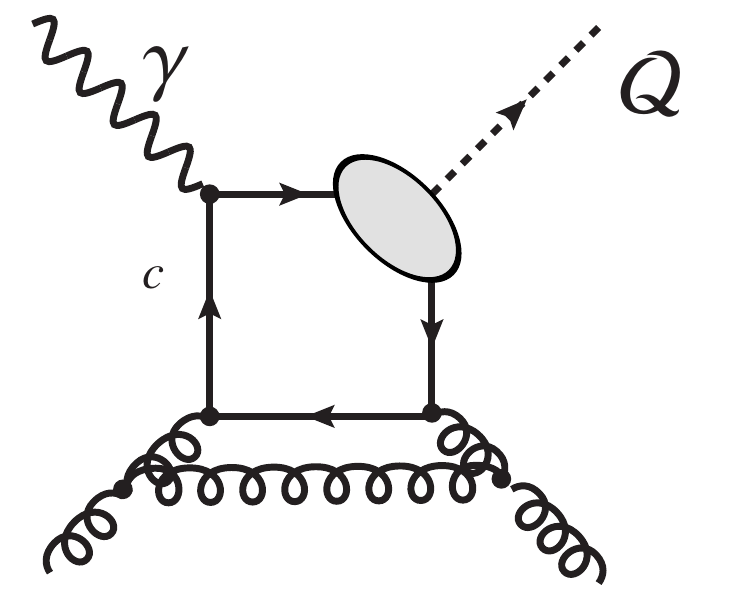}\label{gammag-oniumg-NLO-PT8-CSM}}\\[-4mm]
\subfloat[]{\includegraphics[scale=.25]{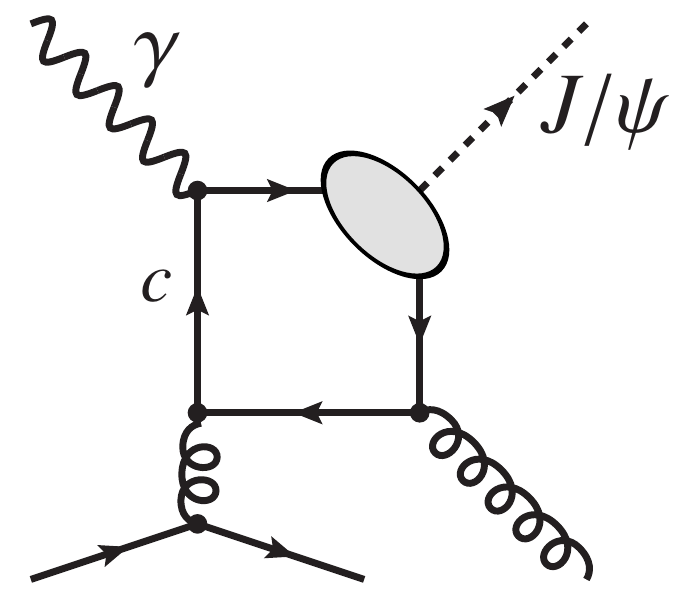}\label{gammaq-psiq-NLO-PT6-CSM}}
\subfloat[]{\includegraphics[scale=.25]{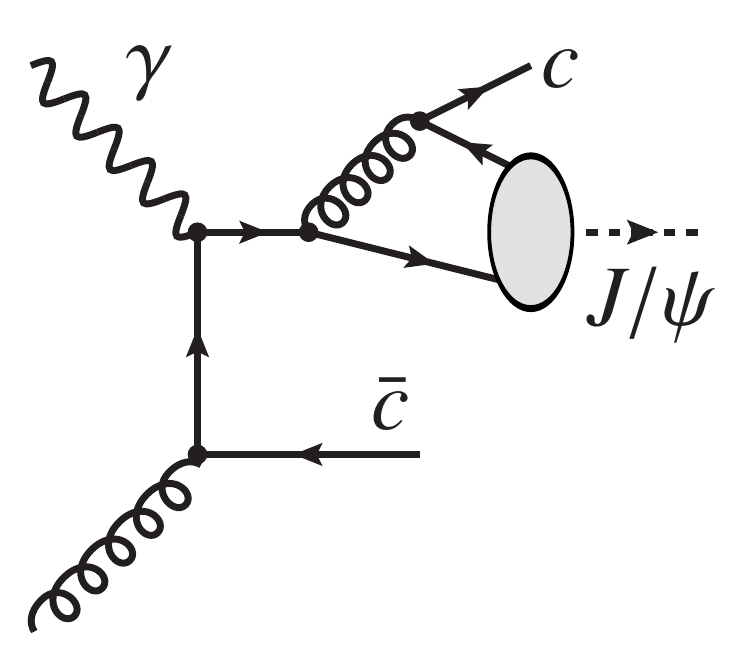}\label{gammag-psicc-CSM}}
\subfloat[]{\includegraphics[scale=.25]{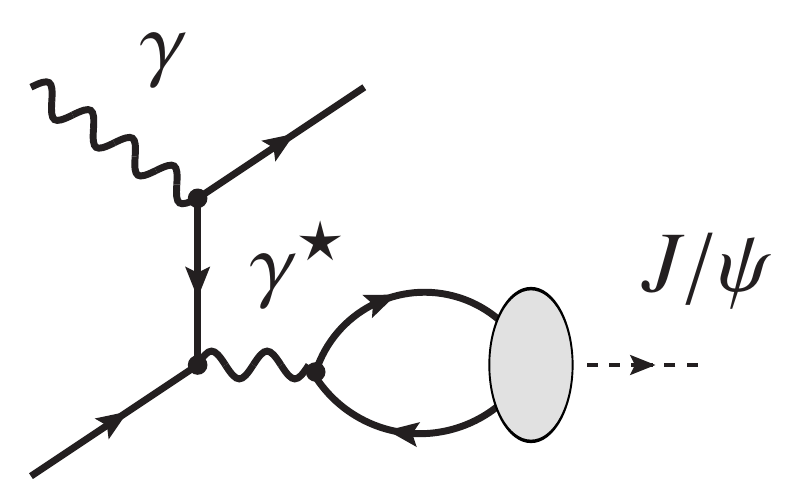}\label{gammaq-psiq-QED-CSM}}
\caption{Representative diagrams contributing to direct $J/\psi$ (unresolved) photoproduction via 
CS channels at orders $\alpha\alpha_s^2$ (a,b), $\alpha\alpha_s^3$ (c,d,e,f) and $\alpha^3$ (g).
The quark and antiquark attached to the ellipsis are taken as on-shell
and their relative velocity $v$ is set to zero.\vspace*{-0.25cm}}
\label{diagrams-CSM-photoproduction}
\end{figure}

At NLO, thus at $\alpha \alpha_s^3$, one additional gluon connected to the above
topologies gives rise to real-gluon-emission contributions, $\gamma+g \to J/\psi+g+g$ (\cf{gammag-psig-NLO-PT6-CSM}) along with loop corrections to
$\gamma+g \to J/\psi+g$ (\cf{gammag-oniumg-NLO-PT8-CSM}). Real-quark-emission contributions also become possible $\gamma+\{q,\bar q\} \to J/\psi+\{q,\bar q\}+g$ (\cf{gammaq-psiq-NLO-PT6-CSM}). Similar radiative corrections to $\gamma+ \{c,\bar c\} \to J/\psi+\{c,\bar c\}$ can be considered. However, the most interesting $\alpha \alpha_s^3$ contribution involving a charm quark in the final state is that of $\gamma + g \to J/\psi + c +\bar{c}$ as it includes the charm fragmentation channels (\cf{gammag-psicc-CSM}).

Even though the contribution of the leading-$P_T$ fragmentation contribution have been discussed in a couple of studies~\cite{Godbole:1995ie,Kniehl:1997gh}, the QED fragmentation contribution (\cf{gammaq-psiq-QED-CSM}), whereby the photon strikes a quark off the proton which then radiates an off-shell photon turning into a $J/\psi$,  has never been studied. It is indeed suppressed by the third power of $\alpha$ but should give a similar contribution than a similar graph with an off-shell radiated gluon into a $^3S_1^{[8]}$ CO. 
As it is expected to scale like $P_T^{-4}$, it can be relevant at large $P_T$. In addition, in the valence region, it may be enhanced by the larger quark densities compared to that of the gluons.

Qiu \etal\ recently reported~\cite{Qiu:2020xum} on the first NLO NRQCD study of the $Q^2$-integrated cross section for $J/\psi$ production at the EIC. Since the yield is expected to be dominated by photoproduction (very low $Q^2$) as opposed to leptoproduction (finite $Q^2$), we find it useful to stress the differences between both their and our studies. In fact, they started with Born-order (LO) contributions from $2 \to 1$ partonic processes, at $\alpha \alpha_s^2$, from CO states. Such LO contributions only lie at $P_T^\star = 0$\footnote{$P_T^\star$ being the transverse momentum measured in the $\gamma^\star p$ center-of-mass system (c.m.s.)}. As such, the corresponding virtual corrections only contribute at $P_T^\star = 0$ and the real-emission corrections generate the $P_T^\star$ spectrum.   
In the photoproduction region, on which we focus here, $P_T^\star \simeq P_T$. Out of their NLO contributions, 
the virtual corrections are thus cut away by the low-$P_T$ cut usually applied to the data (see section~\ref{subsec:ep_formulae}), whereas their real-emission CS corrections
are in fact our $\alpha \alpha_s^2$ $\gamma + g \to J/\psi + g$ contribution.  Going further, 
our $\alpha \alpha_s^3$ $\gamma +\{g, q,\bar q\} \to J/\psi +\{g,q,\bar q\}+g$ contributions would be part of a NNLO analysis of their $Q^2$-integrated observable. 

\subsection{The CSM partonic amplitude}	
\label{subsec:gammaa_formulae}

In the CSM~\cite{Chang:1979nn,Berger:1980ni,Baier:1983va}, one computes the matrix element to create
 a $^3S_1$  quarkonium ${\Q}$ with a momentum $P_\Q$ and a polarisation $\lambda$ accompanied by other partons, commonly noted $\{j\}$, 
by building the product of the amplitude to photoproduce the corresponding heavy-quark pair, ${\cal M}(\gamma+a \to Q \bar Q+\{j\}))$, a spin projector $N(\lambda| s_1,s_2)$ and $R(0)$, the radial wave function at the origin in the configuration space. More precisely, we have
\eqs{\label{eq:CSM_generic}
&{\cal M}(\gamma+a \to {\Q}^\lambda(P_\Q)+\{j\})= \sum_{s_1,s_2,i,i'}\!\!\frac{N(\lambda| s_1,s_2)}{ \sqrt{m_Q}} \frac{\delta^{ii'}}{\sqrt{N_c}} 
 \frac{R(0)}{\sqrt{4 \pi}} \\  \times & 
{\cal M}(\gamma+a \to Q^{s_1}_i \bar Q^{s_2}_{i'}(\mathbf{p}=\mathbf{0})  + \{j\}),}
where $P_\Q=p_Q+p_{\bar Q}$, $p=(p_Q-p_{\bar{Q}})/2$, $s_1$ and $s_2$ are the heavy-quark spins, and $\delta^{ii'}/\sqrt{N_c}$ is the projector onto a CS state. $N(\lambda| s_i,s_{i'})$ is a spin projector. In the non-relativistic limit, $N(\lambda| s_i,s_{i'})= \frac{1}{2 \sqrt{2} m_{Q} } \bar{v} (\frac{\mathbf{P}}{2},s_{i'}) \Gamma_{S} u (\frac{\mathbf{P}}{2},s_i) \,\, $ 
with $\Gamma_S=\varepsilon^{\lambda}_{\mu}\gamma^{\mu}$. When one then sums  over the quark spins, one obtains usual traces which can be evaluated without any specific troubles. In fact, such a computation can be automated at tree level as done by {\sc HELAC-Onia}~\cite{Shao:2012iz, Shao:2015vga}

\subsection{Photoproduction cross section in $ep$ collisions}
\label{subsec:ep_formulae}

Let us start with some elements of kinematics. We first define $s_{ep}=(P_e + P_p)^2=4 E_e E_p$ ($E_{e(p)}$ is the electron (proton) beam energy) and $s_{\gamma p}=W_{\gamma p}^2 = (P_\gamma + P_p)^2$. Introducing $x_\gamma$ as $P_\gamma= x_\gamma P_e$, one has $s_{\gamma p}=x_\gamma s_{ep}$.

Diffractive  contributions are important at low $P_T$ and near the exclusive limit, \ie\ when the $J/\psi$ takes over the entirety of the photon momentum. A cut on $P_{\Q T}$ is usually sufficient to get rid of them. However, 
one can further cut on a variable called elasticity, defined as $z = \tfrac{P_{\Q}\cdot P_p}{P_\gamma \cdot P_p}$. $z$ corresponds to the fraction of the photon energy taken by the $J/\psi$ in the proton rest frame, with the proton momentum defining the z axis. It can equivalently be expressed as $z=\tfrac{2\,E_p \,m_T}{W^2_{\gamma p}\,e^{y}}$ in terms of the $J/\psi$ rapidity $y$ (with $y$ and $E_p$ defined in the same frame) and  the quarkonium transverse mass, $m_T = \sqrt{m_\Q^2 + P_{\Q T}^2}$ . Diffractive contributions usually lie at $z$ close to unity. On the contrary, the resolved-photon contributions can become important at very low $z$ where only a fraction of the photon energy is involved in the scattering.

All our computations are done using {\sc HELAC-Onia} which is well tested for many colliding systems. Let us illustrate now with a generic $2\to 2$ partonic reaction, $\gamma+a \to {\Q}+k$, the connexion between its amplitude, to be computed as discussed in section~\ref{subsec:gammaa_formulae}, and the $ep$ (differential) cross sections which are the measurable quantities. The $ep$ double-differential  cross section in $P_T$ and $z$ is then obtained via a convolution with the proton PDF and the photon flux from the electron :
 \begin{equation}
\begin{aligned}
 \frac{d\sigma}{dzdP_T} & =\int_{x_\gamma^{\rm min}}^1 dx_\gamma \frac{2 x_a P_T f_{\gamma/e}(x_\gamma,Q^2_{max}) f_{a/p}(x_a(x_\gamma),\mu_F)}
{z(1-z)}\\
& \times\frac{1}{16 \pi \hat s^2} \overline{\left| {{\cal M}(\gamma+a \to {\Q}+k)}\right|^2},
\end{aligned}
\end{equation}
\noindent where $x_a=\tfrac{M_T^2-m_{_\Q}^2 z}{x_\gamma s_{e p} z (1-z)}$ and $x_\gamma^{\rm min}= \tfrac{M_T^2-m_{_\Q}^2 z}{s_{e p}z(1-z)}$. 

Following Eq.~(5) of Ref.~\cite{Kniehl:1996we} under the Weizs\"acker-Williams approximation, we have taken 
\begin{equation}
\label{eq:photon-flux}
\begin{aligned}
 &f_{\gamma/e}(x_\gamma,Q^2_{\rm max}) = \frac{\alpha}{2\pi} \times
\\& \Biggl[\frac{1 + (1-x_\gamma)^2}{x_\gamma}\ln\frac{Q^2_{\rm max}}{Q^2_{\rm min}(x_\gamma)}+ 2m_e^2 x_\gamma\lf(\frac{1}{Q^2_{\rm max}} - \frac{1}{Q^2_{\rm min}(x_\gamma)}\rg) \Biggr]\,,\\[1mm]
\end{aligned}
\end{equation}
where $Q^2_{\rm min}(x_\gamma)=m_e^2 x_\gamma^2 / (1-x_\gamma)$ and $m_e$ is the electron mass. 

Besides the specific treatment of the $J/\psi\, +$~charm yield, which we discuss below, the rest of the procedure remains completely standard. For $2\to3$ channels, the formulae would differ but the treatment is also absolutely standard.

\subsection{\label{sec:NLOvalidation}Leading-$P_T$ approximation of the NLO corrections to $J/\psi+g$: NLO$^\star$}

In our evaluation of the NLO corrections to $J/\psi+g$, we will resort to the NLO$^\star$ approximation~\cite{Artoisenet:2008fc,Lansberg:2008gk} which consists in considering the leading-$P_T$ contributions at NLO. These contributions are obtained by imposing a lower cut on the invariant mass of every pair of massless partons, $s_{ij}^{\rm min}$. The \NLOstar approximation avoids the (slow) computations of loop corrections and is easily interfaceable with Monte Carlo (MC) generators, which is advantageous for future EIC simulations.  It has successfully been tested in hadroproduction of single $J/\psi$~\cite{Lansberg:2008gk} and of $J/\psi+J/\psi / \gamma /Z$ pairs~\cite{Lansberg:2009db,Gong:2012ah,Lansberg:2013qka,Lansberg:2014swa}. 

The underlying idea of this approximation is that, for the leading-$P_T$ topologies associated to real NLO emissions, $s_{ij}$ for any $i,j$ pair grows with $P_T$ and the result becomes insensitive to $s_{ij}^{\rm min}$. Only the subleading-$P_T$ contributions --which normally cancel the IR divergences of the loop contributions-- exhibit a $\log(s_{ij}^{\rm min})$ dependence which is {\it de facto} $P_T$-power suppressed.

\begin{figure}[htp!]
\centering
\includegraphics[width=0.9\columnwidth]{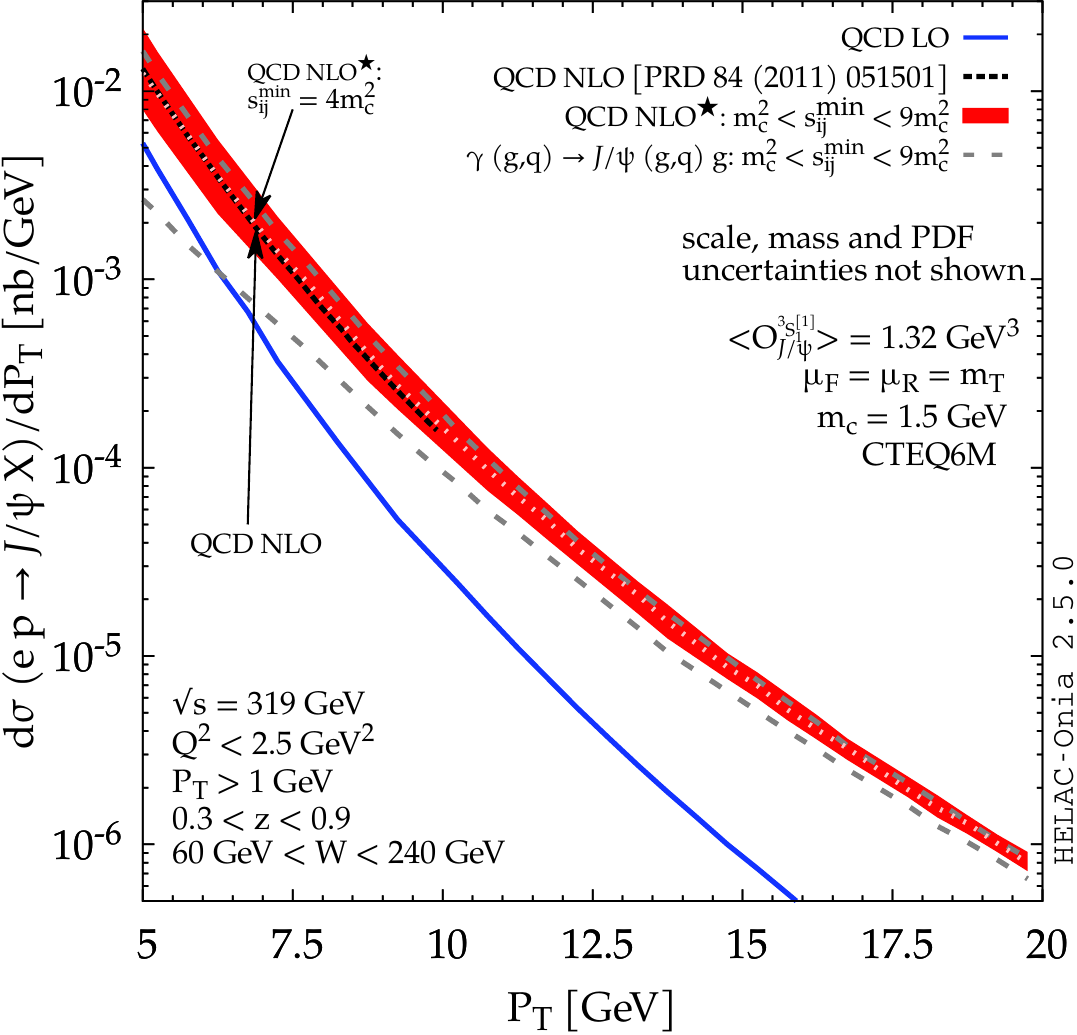}\vspace*{-.25cm}
\caption{Comparison between our \NLOstar (red band) and a complete NLO~\cite{Butenschoen:2011yh} (dashed black line) calculations. The red band results from a variation of the IR cut-off, $s_{ij}^{\rm min}$, specific to the \NLOstar approximation. The dashed grey curves refers to the contributions appearing at $\alpha \alpha_s^3$ for a similar variation of $s_{ij}^{\rm min}$ and the blue at $\alpha \alpha_s^2$ (LO).
}
\label{fig:H1-NLOstar-validation}
\end{figure}

Such an approximation has never been applied to photoproduction whereas the exact same reasoning applies. To check it, we have compared it in Fig.~\ref{fig:H1-NLOstar-validation} (red band), in the HERA Run 2 kinematics for $J/\psi$ photoproduction~\cite{Aaron:2010gz}, to a complete NLO computation by Butensch\"on and Kniehl~\cite{Butenschoen:2011yh} (dashed black line). 
We have used the exact same parameters and PDFs as in Ref.~\cite{Butenschoen:2011yh}: $m_c = 1.5\GeV$, $\mu_F = \mu_R = m_T$, $\langle \mathcal{O}_{J/\psi}^{^{3\!}S_{1}^{[1]}} \rangle = 1.32$\GeV$^3$ and the CTEQ6M proton PDF set~\cite{Pumplin:2002vw}. 
One observes a steady reduction
of the band width from the variation of $\sqrt{s_{ij}^{\rm min}}/m_c$ within the interval $[1,3]$. At $P_T=15$ GeV, the corresponding uncertainty is already less than 30 \% which is smaller than the other uncertainties that we will detail later. Setting $\sqrt{s_{ij}^{\rm min}}/m_c=2$ remarkably reproduces the full NLO result. In what follows, we will use this value for all our \NLOstar predictions.  

One has found out that, for the same $s_{ij}^{\rm min}$ variation at a given $P_T$, the resulting uncertainty is slightly larger than for hadroproduction ($J/\psi+X$~\cite{Lansberg:2008gk} and $J/\psi+\gamma$~\cite{Lansberg:2009db}). This can easily be traced back to the fact that the number of leading-$P_T$ real-emission graphs is twice smaller for photoproduction since there is only a single initial gluon which can radiate. As a case in point, one sees that the LO/NLO ratio is not small at $P_T\sim 5$~GeV when $s_{ij}^{\rm min}/P_T$ is large. In fact, this likely reduces
the  $s_{ij}^{\rm min}$ (lower) uncertainty at $P_T\sim 5$~GeV since the LO does not depend on $s_{ij}^{\rm min}$. This is why we have used a larger variation of $\sqrt{s_{ij}^{\rm min}}/m_c$ for our illustration: $[1,3]$ instead of $[1,2]$ in Refs.~\cite{Lansberg:2008gk,Lansberg:2009db}. For the record, 
the quark channel contributions from $\gamma+q \to J/\psi+g+q$ are at the level of $10\div20 \%$ in this H1 kinematical range.

\subsection{\label{subsec:VFNS}A Variable-Flavour-Number-Scheme treatment of the $J/\psi+$charm yield}

Depending or not whether one considers the presence of charm quarks within the proton, $J/\psi+$charm production
follows either from $\gamma+ g \to J/\psi+c +\bar c$ (3 Flavour Scheme (3FS)) at $\alpha \alpha_s^3$
or $\gamma+ \{c,\bar c\} \to J/\psi+\{c,\bar c\}$ (4 Flavour Scheme (4FS)) at $\alpha \alpha_s^2$.
The situation is similar to single-charm production and both partonic subprocesses can be properly combined using a procedure referred to the Variable Flavour Number scheme (VFNS), outlined in Ref.~\cite{Aivazis:1993pi}. 

A first application to quarkonium production was recently performed by one of us in Ref.~\cite{Shao:2020kgj} to treat $J/\psi+$charm hadroproduction. We go much further here in the photoproduction case considering this channel both from its contribution to the
inclusive yield and to the $J/\psi+$charm yield. We will also study whether a change in the charm quark PDF, motivated by a possible IC component, can be visible in the overall combined cross section.

It is well known that both schemes should yield compatible results, yet they remain complementary in different phase-space regions. The 3FS is better suited close to the heavy-quark threshold with an exact kinematical treatment of the heavy-quark mass 
and the 4FS should give better results when the momenta exchanged are much larger than $m_c$. Indeed, the evolution of the charm PDF
in the 4FS allows one to resum logarithms of ${{\mu_F^2}/{m_c^2}}$, whereas the 3FS only contains one of such logarithms.
Like for bottom-quark-initiated processes~\cite{Maltoni:2012pa}, for increasing scales and $x$, the effect of these initial-state 
logarithms will grow. 

The VFNS allows one to combine both schemes keeping their own virtues without double-counting contributions; it follows 
from the seminal work of Aivazis, Collins, Olness and Tung which lead to the introduction of the ACOT scheme~\cite{Aivazis:1993pi}.
 This has been the object of many studies~\cite{Buza:1996wv,Olness:1997yc,Thorne:1997ga,Cacciari:1998it,Kramer:2000hn,Tung:2001mv,Thorne:2006qt,
Forte:2010ta,Forte:2015hba,Forte:2016sja,Krauss:2017wmx,Forte:2018ovl,Duhr:2020kzd} but none related to quarkonia. 
However, at Born order, the matching procedure is straightforward and easy to implement provided that one properly considers
the issue of the charm-quark mass in the 4FS part given the non-relativistic nature of the $J/\psi$: $m_c$ cannot simply be set to zero.

Rather one should use
\begin{equation}
\begin{aligned}
d\sigma_{\gamma c\to J/\psi+c}&= \frac{1}{2\left(\hat{s}-m_c^2\right)}dx_cf_{c/p}(x_c,\mu_F^2)\\
&\times \overline{\left|\mathcal{M}_{\gamma c\to J/\psi c}\right|^2}d\Phi(p_\gamma,p_c\to P_{\Q},p'_c),
\end{aligned}
\end{equation}
where $d\Phi$ is the final-state phase space, $p_\gamma=\left(\frac{\hat{s}-m_c^2}{2\sqrt{\hat{s}}},0,0,-\frac{\hat{s}-m_c^2}{2\sqrt{\hat{s}}}\right),p_c=\left(\frac{\hat{s}+m_c^2}{2\sqrt{\hat{s}}},0,0,\frac{\hat{s}-m_c^2}{2\sqrt{\hat{s}}}\right)$ in the partonic c.m.s. 
with $\hat{s}=x_cs_{\gamma p}$ since, in the $\gamma p$ c.m.s., $p_{c}=x_{c}P_{p}+\frac{m_c^2}{x_c s_{\gamma p}}P_{\gamma}$.
At LO, the double counting between the 3FS and 4FS contributions can be identified between, on the one hand, 
the initial-gluon splitting into the charm pair in the phase-space region where the charm quark interacting with the photon 
is nearly on-shell and, on the other, the PDF of this charm quark which also comes from a gluon splitting ``inside'' the proton. At LO, the latter reads  
\begin{equation}
f_{c/p}(x_c,\mu_F^2)=\tilde{f}_{c/p}^{(1)}(x_c,\mu_F^2)+\mathcal{O}(\alpha_s^2),
\end{equation}
where
\begin{equation}
\tilde{f}_{c/p}^{(1)}(x_c,\mu_F^2)=\frac{\alpha_s}{2\pi}\log{\left(\frac{\mu_F^2}{m_c^2}\right)}\int_{x_c}^{1}{\frac{dz}{z}P_{qg}(z)f_{g/p}\left(\frac{x_c}{z},\mu_F^2\right)}\label{eq:PDFCT}
\end{equation}
with the well-known Altarelli-Parisi splitting function $P_{qg}(z)=\frac{1}{2}\left[z^2+\left(1-z\right)^2\right]$. The overlap counter-term to be subtracted from the 3FS contribution is then
\begin{equation}
\begin{aligned}
d\sigma_{{\rm CT},\gamma c\to J/\psi+c}&=\frac{1}{2\left(\hat{s}-m_c^2\right)}dx_c\tilde{f}_{c/p}^{(1)}(x_c,\mu_F^2)\\&\times\overline{\left|\mathcal{M}_{\gamma c\to J/\psi c}\right|^2}d\Phi(p_\gamma,p_c\to P_{\Q},p'_c).\label{eq:acCT}
\end{aligned}
\end{equation}
Its computation can be handled by {\sc HELAC-Onia} like the corresponding 4FS contribution.

\vspace*{-0.35cm}
\section{\label{sec:digression-FD} Digression about the feed-down contributions}
\vspace*{-0.25cm}
As mentioned in our introduction, the FD contributions from excited quarkonia or $b$ quarks should be analysed case by case. The ideal situation is when these FD are experimentally determined. For instance, if the $\chi_c$ and $\psi'$ yield are measured, one can deduce more or less directly what fraction of the  $J/\psi$ they  contribute. In the case of inclusive photoproduction, such information is extremely scarce. 

The fraction of non-prompt $J/\psi$ was determined~\cite{Aaron:2010gz} 
for $60 < W_{\gamma p} < 240$~ GeV, 3 bins in $z$ and $P_T> 1$ GeV. In the bin $0.3 < z < 0.4$\footnote{where the resolved-photon contributions are likely negligible.}, it was found to be $14 \pm 7 \%$. No information about its $P_T$ dependence was given but a tuned {\sc Pythia} analysis by H1 pointed at a possibly large value at the highest possible $P_T$ where the inclusive data were taken. Strangely enough, this was left unnoticed until a recent review~\cite{Lansberg:2019adr} by one of us. 

We have found it useful to revisit this by tuning {\sc Pythia 8.2}~\cite{Sjostrand:2014zea} using a $b$ analysis by H1~\cite{Aaron:2012cj} using di-electrons events which extends to large $P_T$. Details are given in~\ref{sec:b-FD}. In agreement with the expectations of H1~\cite{Aaron:2010gz}, we have found the $b$ FD to be significant around $P_T=10$~GeV which probably motivates a dedicated theory analysis at NLO with an updated treatment of the $b\to J/\psi$ hadronisation; this is however beyond the scope of our paper. In our comparison with the H1 data, we will show both the inclusive data (in black) and our expectation for the prompt data (in grey) which has a larger uncertainty but tends to significantly deviate  from the inclusive one.

As what concerns the $\chi_c$ FD, there is currently no theory or experimental indication that it could be relevant and  it is therefore systematically neglected. As for $\psi'$, after the last H1 analysis, ZEUS released in 2012~\cite{Abramowicz:2012dh} the only information which we are aware of, namely the ratio $\psi'/J/\psi$ for 3 points in $P_T$ (up to 3 GeV, though). It was found to be compatible with a LO CS prediction, which directly follows from the ratio of the wave functions at the origin and from the $\psi' \to J/\psi$ branching. In what follows, we will consider it to be on the order of $20\%$.

\vspace*{-0.35cm}
\section{\label{sec:results} Results}
\vspace*{-0.25cm}
In this section, we present and discuss our results for the inclusive prompt yield at HERA and the future EIC. 
At variance with the above comparison betwen the NLO and NLO$^\star$, we use a more modern proton PDF set with uncertainties, namely CT14NLO~\cite{Dulat:2015mca}. We have used the same photon flux as above (see \ce{eq:photon-flux}). The corresponding theoretical uncertainty is evaluated automatically by {\sc HELAC-Onia} using LHAPDF~\cite{Buckley:2014ana}, like the factorisation- and renormalisation-scale uncertainties from an independent variation in the interval $\tfrac{1}{2}\mu_0 < \mu_F, \mu_R < 2\mu_0$, with $\mu_0 = m_T$. We have calculated the mass uncertainty by varying $m_c$ in the range $m_c = 1.5 \pm 0.1\GeV$ for all but the QED channel. Indeed, for this channel, the invariant mass of the photon, $2 m_c$, should coincide with $M_{J/\psi}$ and we chose $m_c = 1.55\GeV$.
We have also taken $\langle \mathcal{O}_{J/\psi}^{^{3\!}S_{1}^{[1]}} \rangle = 1.45$\GeV$^3$ which corresponds\footnote{using $\langle \mathcal{O}_{J/\psi}^{^{3\!}S_{1}^{[1]}} \rangle= \frac{2 (2J+1) N_c\lvert R_{J/\psi}(0)\rvert^2}{ 4 \pi}$.} to
$\lvert R_{J/\psi}(0)\rvert^2 = 1.01\GeV^3$~\cite{Ma:2008gq,Brodsky:2009cf}.

We first naturally start by HERA for which we can compare our computations with existing data. 
Since our study aims at the large(r)-$P_T$ range where the QED and $J/\psi+$~charm contributions can be relevant, we 
will focus on the latest H1 data~\cite{Aaron:2010gz} extending out to $P_T=10$~GeV.
Note that the latest ZEUS measurements~\cite{Chekanov:2002at} only reach  $P_T^{\,max} \simeq 5.5\GeV$.

\begin{figure}[htp!]
\centering 
\includegraphics[width=0.9\columnwidth]{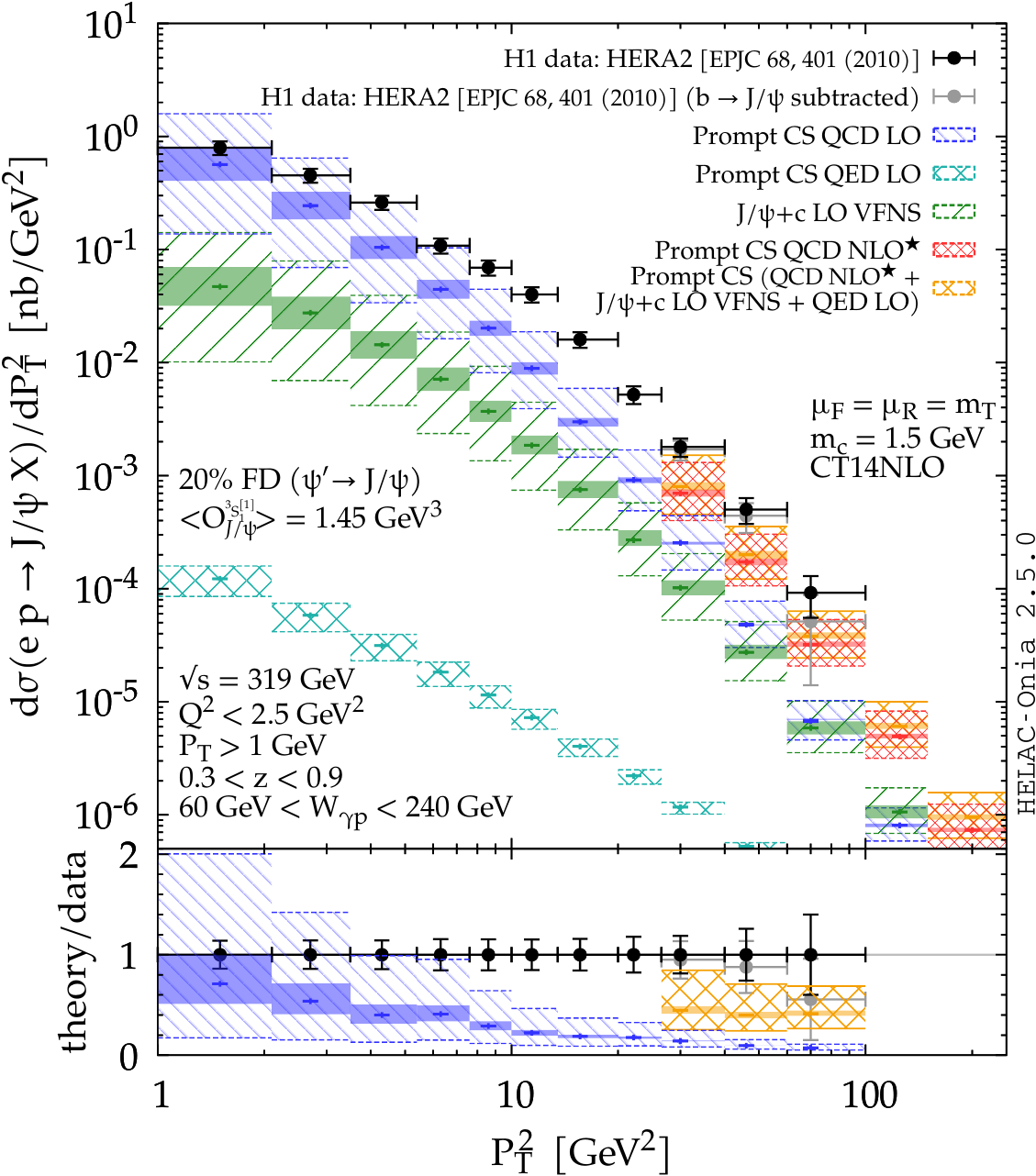}\vspace*{-.25cm}
\caption{Comparison between H1 data~\cite{Aaron:2010gz} (black: inclusive yield; grey: estimated prompt yield) and various CS contributions: LO (blue), \NLOstar (red), LO VNFS  $J/\psi+c$ (green), LO QED (light blue) and their combination (orange). The solid bands indicate the mass uncertainty while the patterns display the scale uncertainty (see text for details).}
\label{fig:H1-CT14NLO-NLOstar}\vspace*{-.25cm}
\end{figure}

\cf{fig:H1-CT14NLO-NLOstar} shows the results for H1 Run 2 kinematics: $\sqrt{s_{ep}} = 319\GeV$, $Q^2 < 2.5\GeV^2$, $P_T > 1\GeV$, $0.3 < z < 0.9$ and $60\GeV < W_{\gamma p} < 240\GeV$. We show our LO (blue) and \NLOstar (red) calculations, the new LO VFNS evaluation of the $J/\psi+$charm contribution (green) to the inclusive yield and the new $\mathcal{O}(\alpha^3)$ QED (light blue) contribution. As discussed in the previous section, the \NLOstar result is plotted for $P_T > 5\GeV$. The sum of all these contributions, which consitutes our best predictions of the CS yield up to $\mathcal{O}(\alpha\alpha_s^2)$ and $\mathcal{O}(\alpha^3)$, is plotted in orange. Its ratio to the data is shown in the lower panel along with that of the LO. For the last 3 bins, we also show our expectation of the prompt yield after the subtraction of the $b$ FD yield (see~\ref{sec:b-FD}). 

About 10 years ago, it was advocated in a couple of works~\cite{Artoisenet:2009xh,Butenschoen:2009zy,Chang:2009uj} that the CSM at NLO was unable to account for the HERA data at large $P_T$ at variance with the interpretation of NLO Kr\"amer's computations~\cite{Kramer:1995nb,Kramer:2001hh}.  We see here that the CSM can indeed describe the data. The introduction of $J/\psi+$charm contribution and the subtraction of the expected $b$ FD improve (compare the red and orange bands) the agreement but the \NLOstar agrees with the data within the uncertainties. We do not wish here to open the debate on whether there is room for additional contribution from CO transition and whether NRQCD NLO fits should be updated with a full study of the CSM uncertainties with modern PDFs and an estimation of $b$ FD. Rather, in view of our discussion for the EIC kinematics, we would like to advocate that a CSM evaluation is certainly valuable in order to provide predictions for future EIC studies in terms of expected rates. In addition, we wish to stress that we anticipate the $J/\psi+$charm contribution to play a more significant role at large $P_T$ and maybe also the QED one in regions where the quark PDF would be less penalised.   

We now move to the future EIC. This collider is expected to reach large integrated luminosities, at different $\sqrt{s_{ep}}$. Here, we focus on two specific configurations, namely those with the following proton-electron beam energies: $100\GeV$ on $5\GeV$ and $275\GeV$ on $18\GeV$. These respectively result  in $\sqrt{s_{ep}} = 45\GeV$ and $\sqrt{s_{ep}} = 140\GeV$.

To avoid the contamination from both diffractive/exclusive and resolved-photon contributions, we apply the following cuts:  $P_T > 1\GeV$ and $0.05 < z < 0.9$.
As also done at HERA, we consider specific ranges in $W_{\gamma p}$. We stress that other choices could be made or that 
$W_{\gamma p}$ could simply be integrated over. In fact, for the lower energy configuration, $\sqrt{s_{ep}} = 45\GeV$, where the cross section is a little smaller, we take a relatively broader interval $10\GeV < W_{\gamma p} < 40\GeV$. For $\sqrt{s_{ep}} = 140\GeV$,
we consider $20\GeV < W_{\gamma p} < 80\GeV$. Finally, we impose a cut on the photon virtuality, $Q^2 < 1\GeV^2$ to remain in the photoproduction region. 

\begin{figure}[ht!]
\centering
\subfloat[$\sqrt{s_{ep}} = 45\GeV$]{\includegraphics[width=0.9\columnwidth]{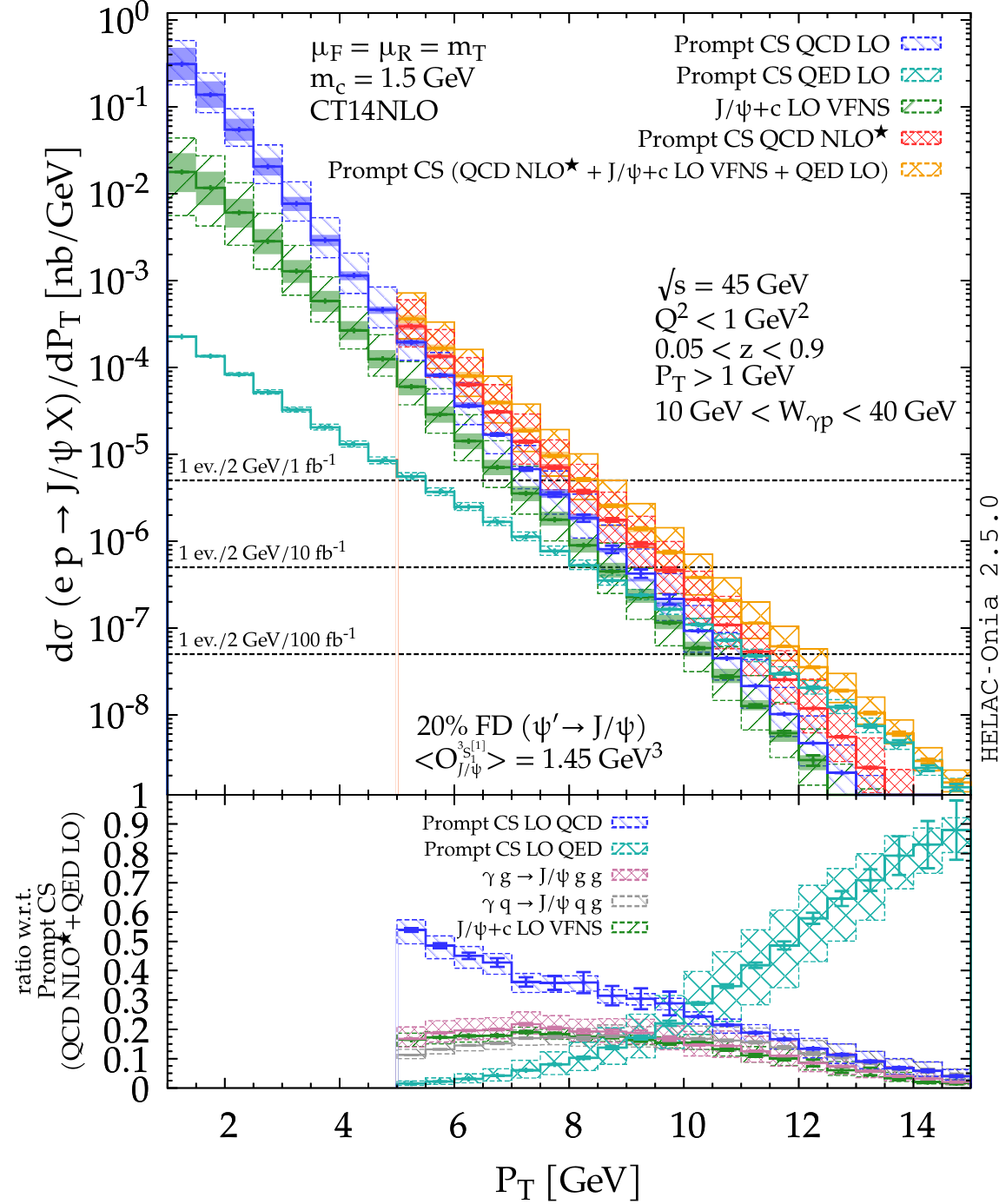}\label{fig:EIC-45GeV-NLOstar}}\vspace*{-.25cm}\\
\subfloat[$\sqrt{s_{ep}} = 140\GeV$]{\includegraphics[width=0.9\columnwidth]{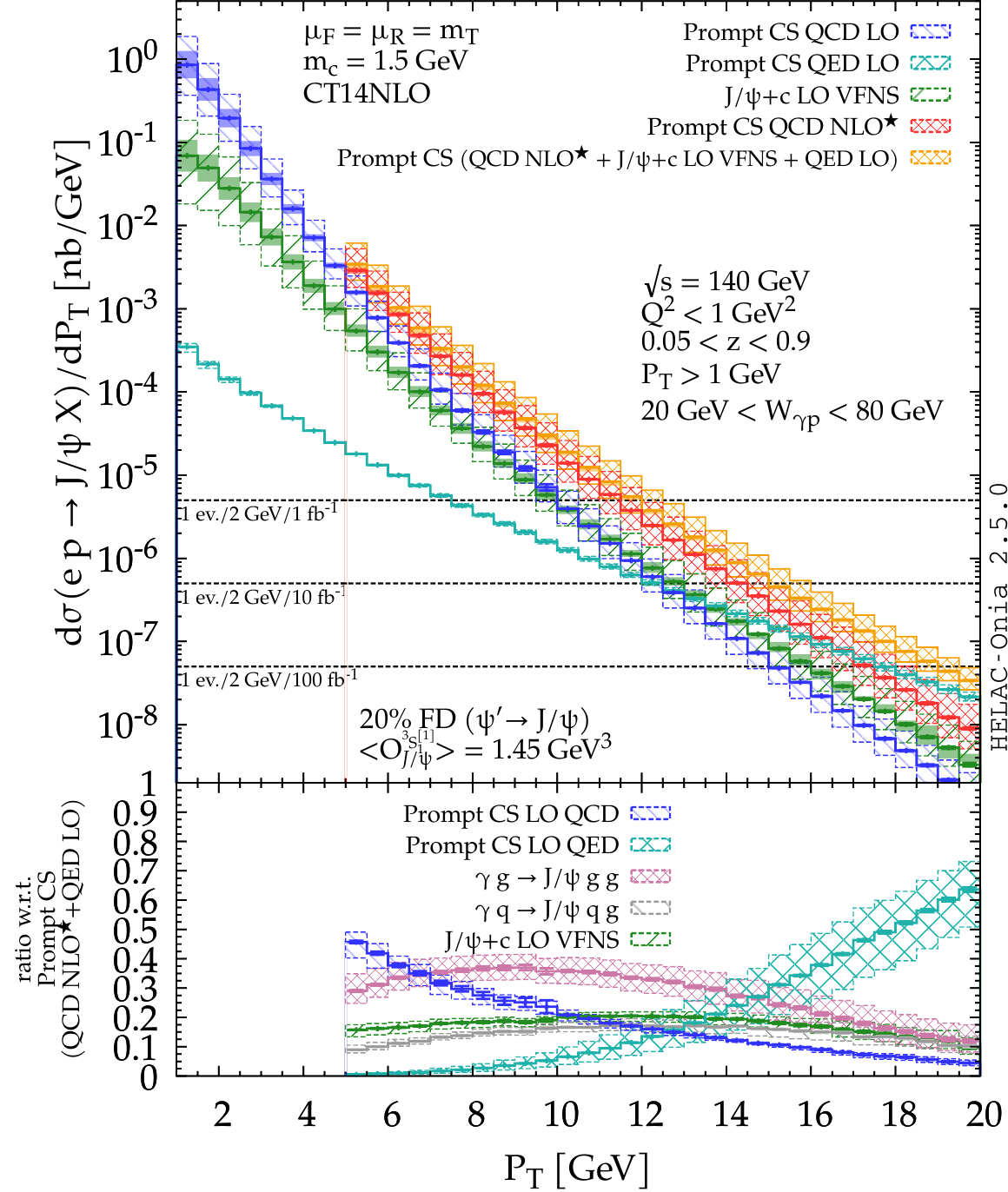}\label{fig:EIC-140GeV-NLOstar}}
\vspace*{-.25cm}
\caption{Predictions for the future EIC at $\sqrt{s_{ep}} = 45$\GeV\ (a) and $\sqrt{s_{ep}} = 140$\GeV\ (b). The calculation is performed adopting the same $\mu_F$, $\mu_R$ and PDFs as \cf{fig:H1-CT14NLO-NLOstar} with the same meaning for bands.}
\label{fig:EIC-CT14NLO-NLOstar}\vspace*{-.25cm}
\end{figure}

Fig.~\ref{fig:EIC-CT14NLO-NLOstar} shows the results for both the $45\GeV$ (a) and $140\GeV$ (b) EIC configurations. The colour code coincides with that of \cf{fig:H1-CT14NLO-NLOstar}. On the lower panel of both plots, the ratio of each contribution to their total is shown.
Let us start by first examining the lowest energy configuration at $\sqrt{s_{ep}} = 45\GeV$. First, we notice that one enters the valence region when $P_T$ increases. Indeed, the $P_T^{-4}$ QED contribution is no more suppressed by the quark PDF and becomes the leading one (and up to about $50\%$ of the full yield) at the largest measurable\footnote{Expectations for the yields associated to a specific luminosity hypothesis are made with the
assumptions of a $10\%$ global efficiency for the $J/\psi$ detection. It is based on the consideration that the $J/\psi$ can be detected in both the di-electron- and di-muon-decay channels with a high efficiency in the respective EIC detector acceptance, like with LHCb at the LHC.}  $P_T \simeq 10\GeV$ with an integrated luminosity of $\mathcal{L} = 100\,\text{fb}^{-1}$. Second, we note that the $J/\psi +$~unidentified charm contribution is comparable to the $\mathcal{O}(\alpha \alpha_s^3)$ $\gamma g(q)$ fusion subprocesses. In such an EIC configuration, both new photoproduction contributions that were so far overlooked are thus absolutely relevant.

Moving now to the largest energy configuration at $\sqrt{s_{ep}} = 140\GeV$, one notes that the $P_T$ spectrum should be observable up to $P_T\!\sim\!15\div20\GeV$ with $\mathcal{L} = 100\,\text{fb}^{-1}$. Again, the $\mathcal{O}(\alpha^3)$ QED contribution becomes dominant at large $P_T$, being for instance more than $40-50\%$ of the full yield at $P_T \gtrsim 15 \GeV$. At intermediate $P_T$, the photon-gluon fusion prevails and thus $J/\psi$ photoproduction in principle remains a good glue probe there. More generally, the production of $J/\psi + 2$ hard partons is dominant for $10 \lesssim P_T \lesssim 15\GeV$. Such a process could be identified by looking at $J/\psi + 2$ jets of moderate $P_T$. Moreover, one of them should be relatively close to the $J/\psi$ while the other should recoil on both. This clearly motivates dedicated studies to assess the feasibility of such $J/\psi +2 $ jet measurements. For a generic discussion about $J/\psi\, + $ jets production, we guide the reader to~\cite{Lansberg:2019adr}.

\vspace*{-0.35cm}
\section{\label{subsec:charm} $J/\psi\, +$ (identified) charm}
\vspace*{-0.25cm}
We finally move to the associated production of a $J/\psi$ and a charm quark,  leading to the observation of a charmed particle. In the past, unexpectedly large yields for the production of $J/\psi$ and a $c \bar c$ pair in $e^+e^-$ annihilation were measured by the Belle~\cite{Abe:2002rb,Pakhlov:2009nj} collaboration at KEK. Such large yields could finally be explained within the CSM after the inclusion of $\alpha_s^3$ (NLO) corrections~\cite{Zhang:2006ay,Gong:2009ng} with a $K$ factors on the order of $1.4 \div 1.7$. We guide the reader to Ref.~\cite{Lansberg:2019adr} for a discussion of the different existing theoretical studies.

This naturally motivated  theoretial studies of the corresponding process in the hadroproduction case~\cite{Baranov:2006dh,Artoisenet:2007xi,Brodsky:2009cf,Lansberg:2010vq,Shao:2020kgj}. We extend these here to the photoproduction case with the potential of the future EIC in mind 
and use the same LO VFNS computation as for $J/\psi\,+$ (unobserved) charm channels contributing to the inclusive $J/\psi$ yield which we just discussed. At variance with the results presented in Section~\ref{sec:results}, we now consider a charm tagging efficiency, which we dub $\varepsilon_c$ and which then enters the combination of the 3FS and 4FS contributions (see Section~\ref{subsec:VFNS}) meant to avoid any double counting:
\begin{equation}\label{eq:VFNS_with_efficiency}
 d\sigma^{\rm VFNS} = d\sigma^{\rm 3FS}\left[1-\left(1-\varepsilon_c\right)^2\right] + \left(d\sigma^{\rm 4FS} - d\sigma^{\rm CT}\right)\varepsilon_c.
\end{equation}

Based on EIC simulations~\cite{Chudakov:2016otl,Furletova:INT}, we take $\varepsilon_c = 10\%$. It is a first working hypothesis which includes possibilities to tag the (anti)charm via charmed hadrons or by tagging a charm jet.  In order to also study the possible impact of IC, we adopt the CT14NNLO proton PDF set~\cite{Hou:2017khm}, which includes PDF eigensets with some IC effects, namely a ``sea-like'' one (in green) and a valence-like one, also called ``BHPS'', (in red). The central eigenset include ``no IC'' (in blue). We stress that this only affects the combined cross section via the 4FS contribution through \ce{eq:VFNS_with_efficiency}\footnote{except for some possible effects on the gluon PDF via the momentum sum rule.}.

\cf{fig:EIC-VFNS-IC} shows the results for the $J/\psi$+charm associated production at the EIC at $\sqrt{s_{ep}} = 45\GeV$ (a) and $\sqrt{s_{ep}} = 140\GeV$ (b). On the lower panel of both plots, the ratio to the ``no IC'' case is shown and compared to the ``no IC'' relative uncertainty.
One can first note that at $\sqrt{s_{ep}} = 45\GeV$, even with $\mathcal{L} = 100\,\text{fb}^{-1}$, the yield is limited to $P_T$ below 5 GeV. On the contrary, the $P_T$ reach extends up to $P_T \sim 10-12\GeV$ at $\sqrt{s_{ep}} = 140\GeV$. 
In the former configuration, the BHPS valence-like (in red) enhancement is visible, being as large as $4-5$ times the ``no IC'' yield at $P_T \sim 4-5\GeV$. More importantly, this deviation is larger than the expected uncertainty (also in blue in the lower panel) of the ``no IC'' yield. At $\sqrt{s_{ep}} = 140\GeV$, one does not probe the valence region and such a deviation becomes too small compared to the ``no IC'' uncertainty.

The EIC at $\sqrt{s_{ep}} = 45\GeV$ will thus be the place to probe the non-perturbative charm content of the proton with $J/\psi+$charm production. Besides, it would also be interesting to know how precisely the charm and anticharm contributions to this process could be experimentally separated out in order to look for an IC induced $c(x)$ vs $\bar c(x)$ asymmetry as recently found from lattice QCD studies~\cite{Sufian:2020coz}.
A further motivation to study this process is to investigate on colour transfers~\cite{Nayak:2007mb}, that can enhance the cross section at low  invariant masses of the $J/\psi+$charm pair, $M_{J/\psi+c}$. Such a phenomenon should be both measured in photoproduction and hadroproduction.

\begin{figure}[htb!]
\centering
\subfloat[$\sqrt{s_{ep}} = 45\GeV$]{\includegraphics[width=0.9\columnwidth]{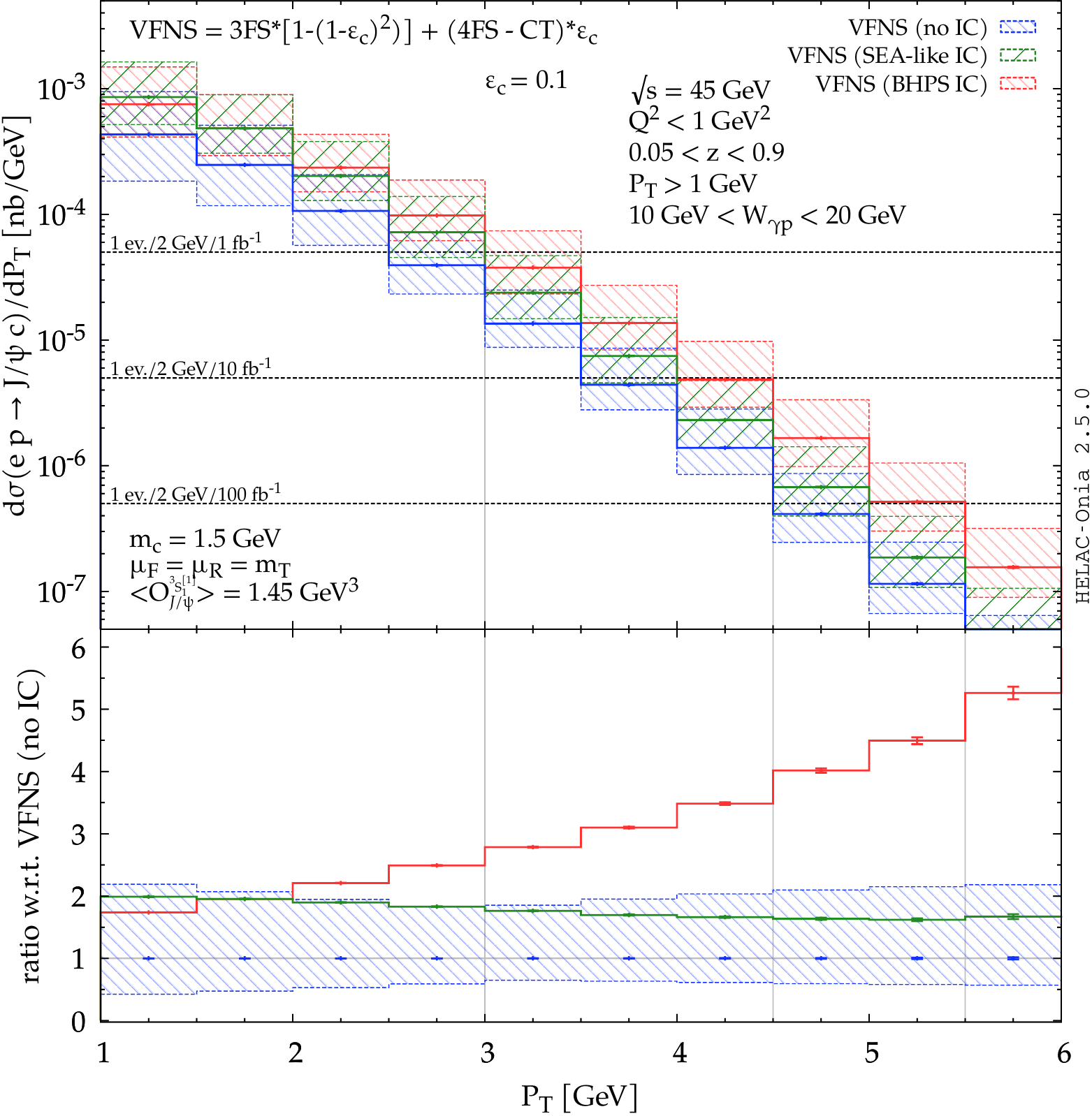}}\\
\subfloat[$\sqrt{s_{ep}} = 140\GeV$]{\includegraphics[width=0.9\columnwidth]{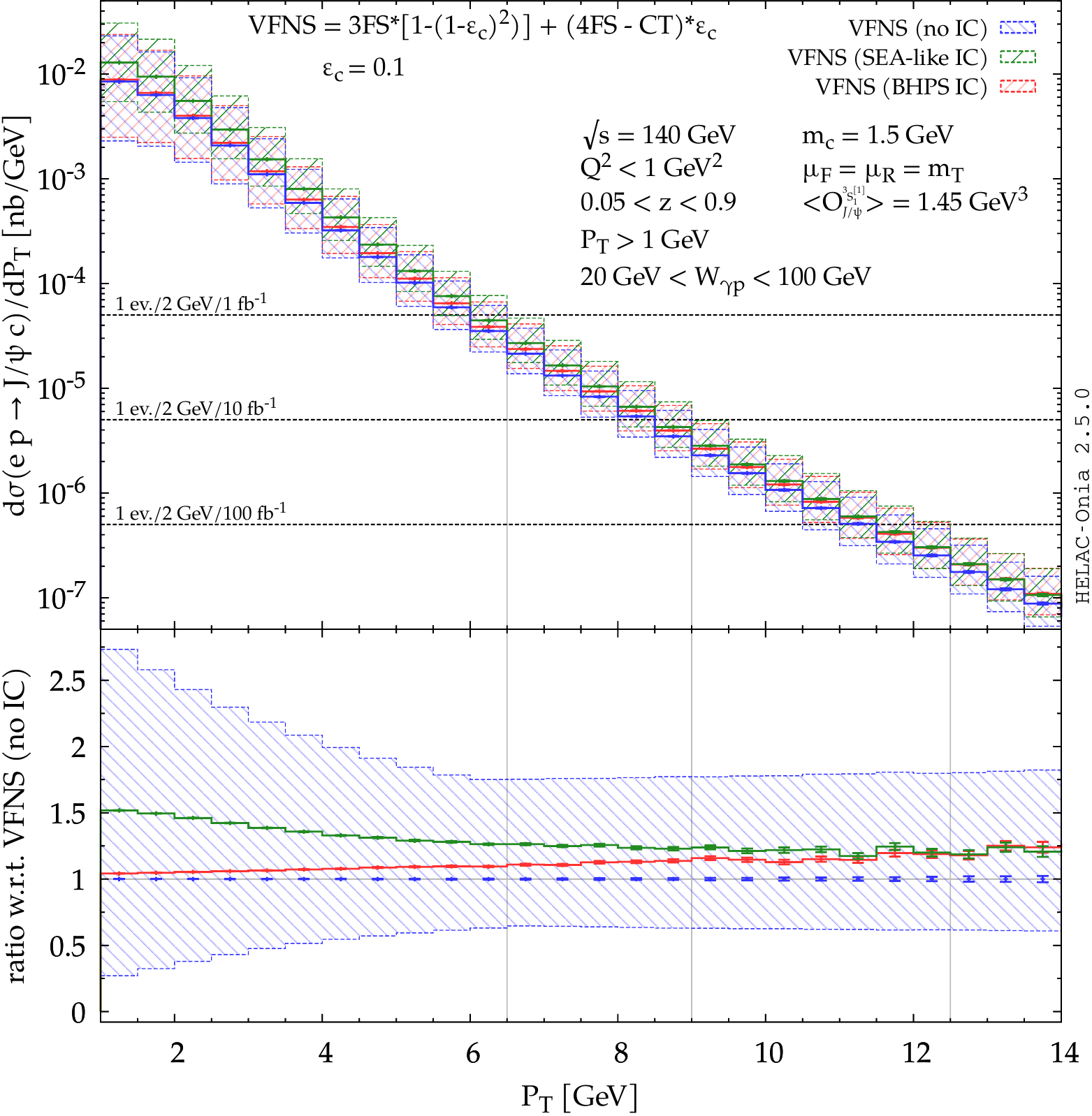}}
\caption{Predictions for $J/\psi +$charm production at the future EIC at $\sqrt{s_{ep}} = 45\GeV$ (a) and $\sqrt{s_{ep}} = 140\GeV$ (b) computed with charm PDFs with ``no IC'' (blue), sea-like (green) and ``BHPS'' valence-like (red) ICs. The lower panel shows the ratio to the ``no IC'' curves and its relative uncertainty. The kinematical cuts are the same as in~\cf{fig:EIC-CT14NLO-NLOstar}. A $10\%$ charm detection efficiency is considered when drawing the horizontal observability lines.}\vspace*{-.25cm}
\label{fig:EIC-VFNS-IC}\vspace*{-.25cm}
\end{figure}

\vspace*{-0.25cm}
\section{\label{sec:conclusions}Conclusions}
\vspace*{-0.20cm}

In this Letter, we have analysed the inclusive, large-$P_T$ $J/\psi$ photoproduction at $ep$ colliders. Such a process is a key measurement to improve our understanding of the production mechanism of quarkonia and a potential good probe of gluons. 
Our analysis incorporates the leading-$P_T$ leading-$v$ next-to-leading $\alpha_s$ corrections, \ie~the \NLOstar CSM. 
In addition, we have studied for the first time the $\mathcal{O}(\alpha^3)$ QED and the $J/\psi$+charm contributions, which both also contain leading-$P_T$ topologies. We have computed the $J/\psi$+charm cross section at LO in the VFNS, which properly combines the $\mathcal{O}(\alpha \alpha_s^3)$ 3FS and $\mathcal{O}(\alpha \alpha_s^2)$ 4FS contributions with specific counter-terms, avoiding double counting.

By comparing our \NLOstar cross section to a complete NLO calculation at HERA, we have further validated this leading-$P_T$ approximation, already used in hadroproduction for a few processes. We have then reassessed the latest H1 data. It turns out that, contrary to earlier claims, the CSM incorporating NLO leading-$P_T$ contributions can describe $J/\psi$ photoproduction at HERA for $P_T > 5\GeV$. 
The agreement with the data slightly improves when the $J/\psi+$charm contribution is included, while the LO QED contribution would only play a role at larger $P_T$ (on the order of 20 GeV).
We have also confirmed with a simple data-driven analysis that the $b$ FD should probably not be ignored as it has so far been the case.
When approximately subtracted from the data, the data-CSM agreement indeed further improves.

We have then studied the case of a future EIC which will be able to run at high luminosity (up to $100\,\text{fb}^{-1}$ for one year of data taking) with different $\sqrt{s_{ep}}$ energies. As such, the EIC should have a larger $P_T$ reach. We have provided predictions for the inclusive cross section for prompt $J/\psi$ photoproduction at both the lowest and highest $\sqrt{s_{ep}}$ configurations, namely $45\GeV$ and $140\GeV$.
At variance with HERA, the $P_T^{-4}$ QED contribution turns out to be dominant at large $P_T$, becoming about half of the prompt yield at $P_T \simeq 10\,(15)\GeV$ for the $\sqrt{s_{ep}} = 45\,(140)\GeV$ configuration. This is due to the prominent role of the quark in the valence region. 
Furthermore, we have found that the $J/\psi+c$ contribution is on the same order as the $\mathcal{O}(\alpha \alpha_s^3)$ $\gamma g(q)$ subprocesses at $\sqrt{s_{ep}} = 45\GeV$. It is thus important to note that $J/\psi$ production at the lower energy configuration of the EIC is no longer a gluon probe at mid and large $P_T$. The $\mathcal{O}(\alpha \alpha_s^3)$ $\gamma g(q)$ processes
can however be disentangled by looking at a  $J/\psi$ with 2 jets of moderate $P_T$.

Applying the VFNS to the $J/\psi+$charm production, we have also demonstrated its observability at the EIC. 
At $\sqrt{s_{ep}} = 45\GeV$, a ``BHPS'' valence-like be\-hav\-iour can be observed at $P_T > 5\GeV$, with a possible enhancement as large as $4-5$ compared to a ``no IC'' hypothesis . Indeed, in this energy configuration this processes is sensitive to the charm quark in the valence region. This is not so at $\sqrt{s_{ep}} = 140\GeV$, where one cannot a priori discriminate between the ``IC'' and ``no IC'' hypotheses.
The EIC at $\sqrt{s_{ep}} = 45\GeV$ thus offers a unique opportunity to better constrain the charm quark content via the associated production of a $J/\psi$ with an identified charmed particle.

In the future, such studies could surely be  extended to other planned colliders, such as the LHeC and FCC-eh~\cite{AbelleiraFernandez:2012cc, Agostini:2020fmq}. In addition, dedicated experimential studies for the associated production of $J/\psi$+jets and FD from heavier quarks or quarkonia at the EIC would be surely beneficial.

\vspace*{-0.25cm}
\section*{Acknowledgements}
We thank Y.~Feng, M.A.~Ozcelik, J.W.~Qiu, I. Schienbein, H. Spiesberger for useful discussions. 
This project has received funding from the European Union's Horizon 2020 research and innovation programme under grant agreement No.~824093 in order
to contribute to the EU Virtual Access {\sc NLOAccess}.
This work  was also partly supported by the French CNRS via the IN2P3 project GLUE@NLO and via the Franco-Chinese LIA FCPPL (Quarkonium4AFTER), by the Paris-Saclay U. via the P2I Department and by the P2IO Labex via the Gluodynamics project.
H.S.S.~is supported by the ILP Labex (ANR-11-IDEX-0004-02, ANR-10-LABX-63).
L.Y.~is supported by the EU Erasmus+ Paris-Saclay U.-Ukraine program.

\vspace*{-0.25cm}
\appendix
\renewcommand{\thefigure}{\arabic{figure}}
\section{\label{sec:b-FD} Feed-down from b}

In this appendix, we briefly explain the data-driven procedure we have used to estimate the $b\to J/\psi$ FD at HERA.
We have considered the latest H1 analysis on $b \bar b$ production~\cite{Aaron:2012cj} via di-electron events. The $b$ photoproduction cross section was measured in four $\langle P_T(b)\rangle$ bins, with $\langle P_T(b)\rangle = \sqrt{\left(P_{T,b}^2 + P_{T,\bar b}^2\right)/2}$, in the range $[0, 30]\GeV$. It is important to note that H1 reported on a $b$, not on a $B$ cross section.
We have thus computed the corresponding LO plus parton shower (LO+PS) cross section using {\sc Pythia 8.2}, adopting the CT14NLO PDF set 
with  $m_b = 4.75\GeV$ and $\mu_F = \mu_R = m_T$. Note that we have not considered any QCD uncertainty since we have assumed that they are fully correlated between both the H1 kinematics of the $b$ measurements and that of the $J/\psi$ measurements. The corresponding kinematical ranges are indeed very similar. Such theory uncertainties will thus be absorbed in the tuning factor which we discuss now.
We have indeed performed a $\chi^2$-minimisation to compute a tuning factor, denoted $N_{FD}$, such that the obtained LO+PS {\sc Pythia} spectrum reproduce best the H1 $b$ data. We have found $N_{FD} = 3.8 \pm 0.5$. We note at this stage that $N_{FD}$ partially takes into account higher-order effects missing in the LO+PS computation. The resulting cross section is shown in~\cf{fig:H1-Ag2bbbar} (a)  with the H1 values.

This tuning of the $b$ cross section being  made, we have again used  {\sc Pythia 8.2} to compute the $b \to J/\psi$ cross section in the H1 kinematics~\cite{Aaron:2010gz}. Indeed, we expect {\sc Pythia 8.2} to account reasonably well for the fragmentation of a $b$ quark into a $J/\psi$. The obtained cross section is plot in red in~\cf{fig:H1-Ag2bbbar} (b). Its uncertainty is purely from that of $N_{FD}$, thus from the H1 uncertainty of the reported H1 $b$ cross section. As explained above, we have assumed that the theory uncertainties are absorbed in the tuning factor. We have then subtracted this $b \to J/\psi$ yield from the inclusive one 
 in order to obtain a data-driven estimate of the prompt cross section. These are respectively shown in~\cf{fig:H1-Ag2bbbar} (b) \& \cf{fig:H1-CT14NLO-NLOstar}. (black and grey points). To conclude, we note that our data-driven procedure yields a fraction of non-prompt $J/\psi$ slightly smaller, yet compatible with the $P_T$-integrated value reported by H1~\cite{Aaron:2010gz} for $60 < W_{\gamma p} < 240$~ GeV  for $0.3 < z < 0.4$. This means that our data-driven estimate can be seen as a conservative {\it low} value of the $b$ FD to be subtracted at large $P_T$.

\begin{figure}[h!]
\centering
\subfloat[]{\includegraphics[width=8.3cm, keepaspectratio]{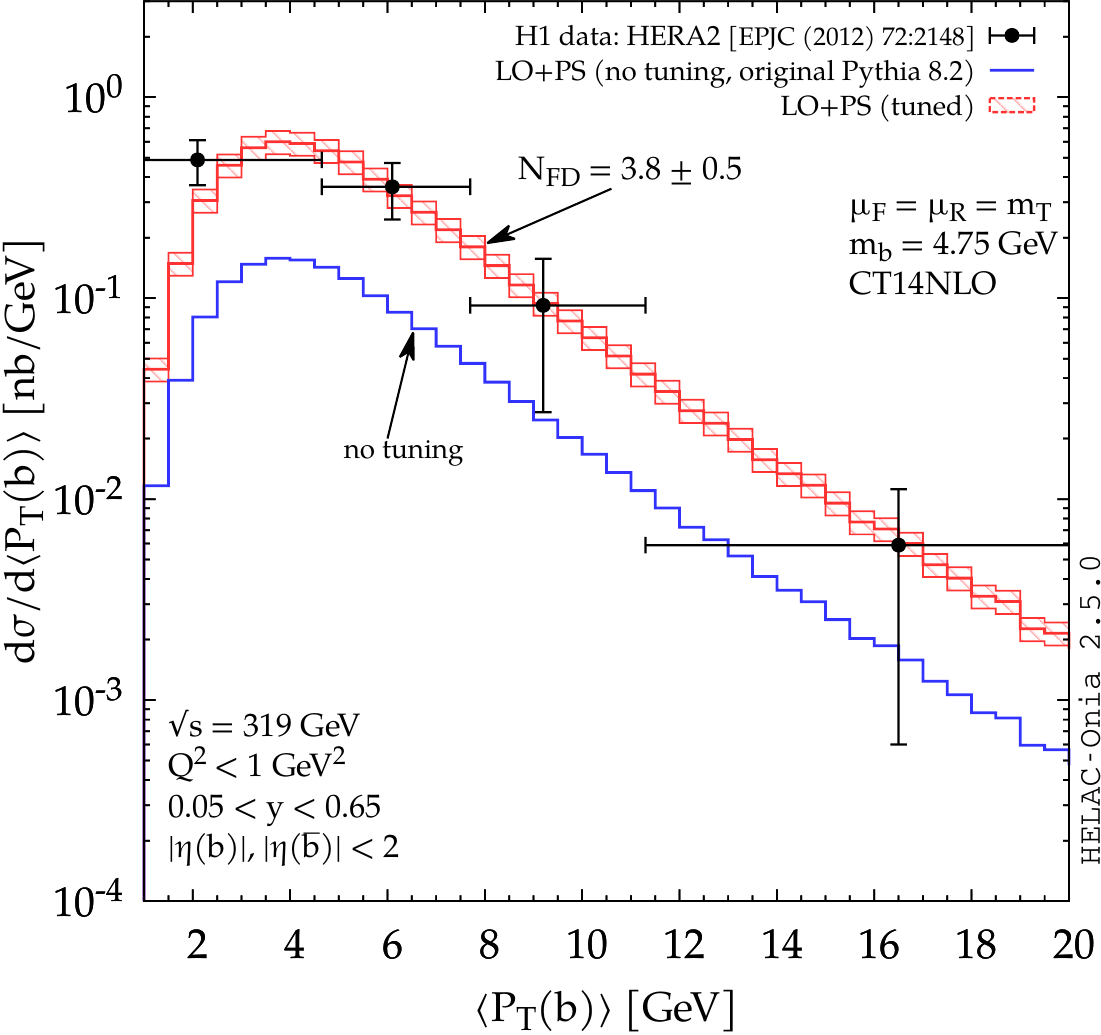}}\vspace*{-.5cm}
\subfloat[]{\includegraphics[width=8.3cm, keepaspectratio]{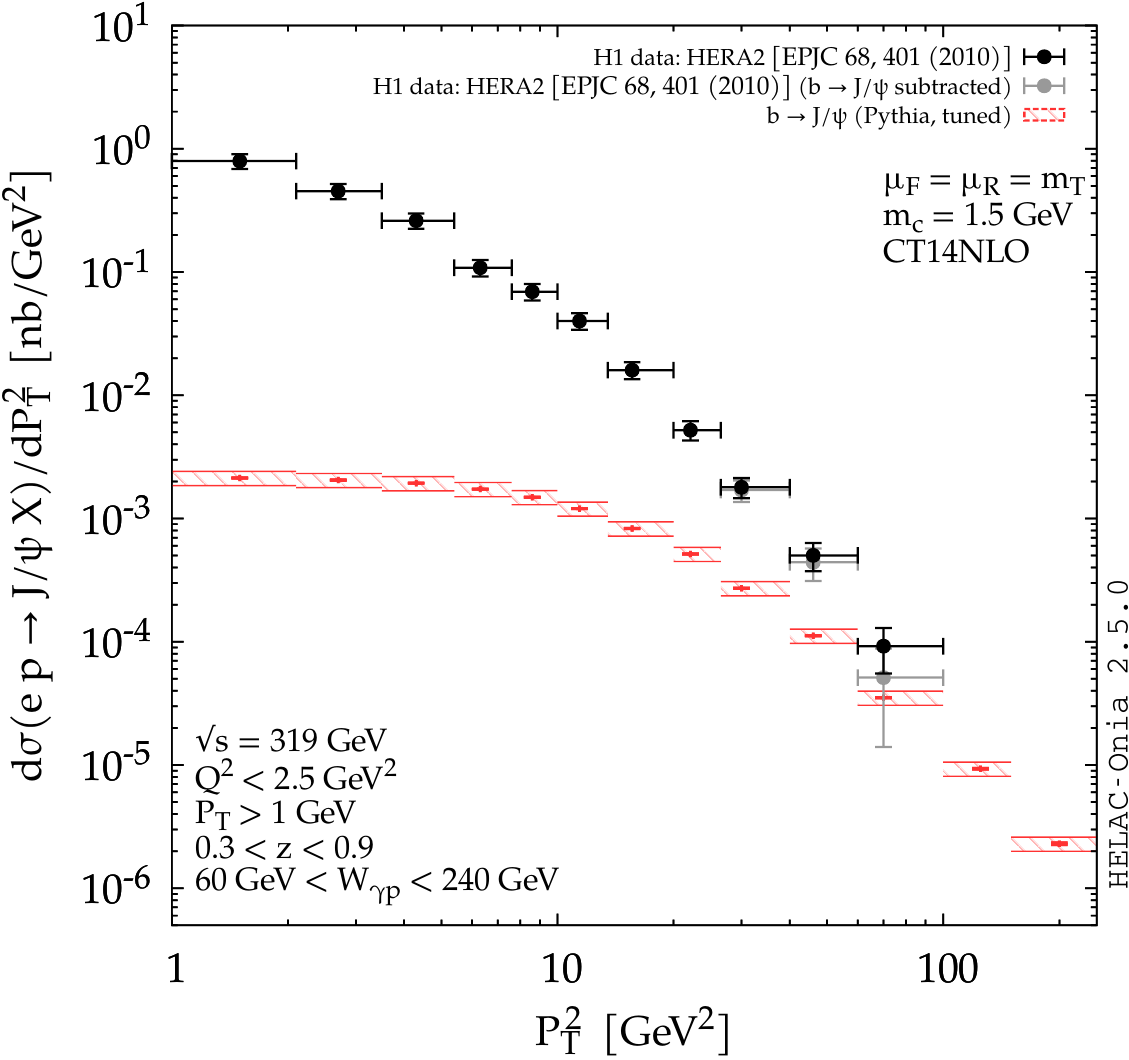}}
\vspace*{-.25cm}
\caption{(a) Comparison between the H1 $b$ cross section~\cite{Aaron:2012cj} and our LO+PS computation. The blue (resp. red) 
histograms depicts the ``untuned'' (resp. tuned) results. The dotted histograms indicates the uncertainty from 
the $\chi^2$-minimisation via $N_{FD} = 3.8 \pm 0.5$.
(b) The measured inclusive (black points) and derived prompt (grey points) H1 $J/\psi$ cross sections~\cite{Aaron:2010gz} along with the 
derived $b \to J/\psi$ one (red) obtained after tuning {\sc Pythia} (red).}
\label{fig:H1-Ag2bbbar}
\end{figure}

\bibliographystyle{utphys}
\biboptions{sort&compress}
\bibliography{high_pT_psi-HERA-EIC-090920}

\providecommand{\href}[2]{#2}\begingroup\raggedright\begin{thebibliography}{10}

\bibitem{Kramer:2001hh}
M.~Kraemer, ``{Quarkonium production at high-energy colliders},''
  \href{http://dx.doi.org/10.1016/S0146-6410(01)00154-5}{{\em Prog. Part. Nucl.
  Phys.} {\bfseries 47} (2001) 141--201},
\href{http://arxiv.org/abs/hep-ph/0106120}{{\ttfamily arXiv:hep-ph/0106120
  [hep-ph]}}.

\bibitem{Brambilla:2004wf}
{\bfseries Quarkonium Working Group} Collaboration, N.~Brambilla {\em et~al.},
  ``{Heavy quarkonium physics},''
\href{http://arxiv.org/abs/hep-ph/0412158}{{\ttfamily arXiv:hep-ph/0412158
  [hep-ph]}}.

\bibitem{Lansberg:2006dh}
J.~P. Lansberg, ``{$J/\psi$, $\psi'$ and $\Upsilon$ production at hadron
  colliders: A Review},''
  \href{http://dx.doi.org/10.1142/S0217751X06033180}{{\em Int. J. Mod. Phys.}
  {\bfseries A21} (2006) 3857--3916},
\href{http://arxiv.org/abs/hep-ph/0602091}{{\ttfamily arXiv:hep-ph/0602091
  [hep-ph]}}.

\bibitem{Brambilla:2010cs}
N.~Brambilla {\em et~al.}, ``{Heavy Quarkonium: Progress, Puzzles, and
  Opportunities},''
  \href{http://dx.doi.org/10.1140/epjc/s10052-010-1534-9}{{\em Eur. Phys. J.}
  {\bfseries C71} (2011) 1534},
\href{http://arxiv.org/abs/1010.5827}{{\ttfamily arXiv:1010.5827 [hep-ph]}}.

\bibitem{Andronic:2015wma}
A.~Andronic {\em et~al.}, ``{Heavy-flavour and quarkonium production in the LHC
  era: from proton–proton to heavy-ion collisions},''
  \href{http://dx.doi.org/10.1140/epjc/s10052-015-3819-5}{{\em Eur. Phys. J.}
  {\bfseries C76} no.~3, (2016) 107},
\href{http://arxiv.org/abs/1506.03981}{{\ttfamily arXiv:1506.03981 [nucl-ex]}}.

\bibitem{Lansberg:2019adr}
J.-P. Lansberg, ``{New Observables in Inclusive Production of Quarkonia},''
  {\em Phys.Rept.} (2020) , \href{http://arxiv.org/abs/1903.09185}{{\ttfamily
  arXiv:1903.09185 [hep-ph]}}.
(In press).

\bibitem{Aid:1996dn}
{\bfseries H1} Collaboration, S.~Aid {\em et~al.}, ``{Elastic and inelastic
  photoproduction of $J/\psi$ mesons at HERA},''
  \href{http://dx.doi.org/10.1016/0550-3213(96)00274-X}{{\em Nucl. Phys. B}
  {\bfseries 472} (1996) 3--31},
  \href{http://arxiv.org/abs/hep-ex/9603005}{{\ttfamily arXiv:hep-ex/9603005}}.

\bibitem{Breitweg:1997we}
{\bfseries ZEUS} Collaboration, J.~Breitweg {\em et~al.}, ``{Measurement of
  inelastic $J/\psi$ photoproduction at HERA},''
  \href{http://dx.doi.org/10.1007/s002880050583}{{\em Z. Phys. C} {\bfseries
  76} (1997) 599--612}, \href{http://arxiv.org/abs/hep-ex/9708010}{{\ttfamily
  arXiv:hep-ex/9708010}}.

\bibitem{Chekanov:2002at}
{\bfseries ZEUS} Collaboration, S.~Chekanov {\em et~al.}, ``{Measurements of
  inelastic $J /\psi$ and $\psi^\prime$ photoproduction at HERA},''
  \href{http://dx.doi.org/10.1140/epjc/s2002-01130-2}{{\em Eur. Phys. J. C}
  {\bfseries 27} (2003) 173--188},
  \href{http://arxiv.org/abs/hep-ex/0211011}{{\ttfamily arXiv:hep-ex/0211011}}.

\bibitem{Adloff:2002ex}
{\bfseries H1} Collaboration, C.~Adloff {\em et~al.}, ``{Inelastic
  photoproduction of $J/\psi$ mesons at HERA},''
  \href{http://dx.doi.org/10.1007/s10052-002-1009-8}{{\em Eur. Phys. J.}
  {\bfseries C25} (2002) 25--39},
\href{http://arxiv.org/abs/hep-ex/0205064}{{\ttfamily arXiv:hep-ex/0205064
  [hep-ex]}}.

\bibitem{Chekanov:2009ad}
{\bfseries ZEUS} Collaboration, S.~Chekanov {\em et~al.}, ``{Measurement of
  J/psi helicity distributions in inelastic photoproduction at HERA},''
  \href{http://dx.doi.org/10.1088/1126-6708/2009/12/007}{{\em JHEP} {\bfseries
  12} (2009) 007},
\href{http://arxiv.org/abs/0906.1424}{{\ttfamily arXiv:0906.1424 [hep-ex]}}.

\bibitem{Aaron:2010gz}
{\bfseries H1} Collaboration, F.~Aaron {\em et~al.}, ``{Inelastic Production of
  $J/\psi$ Mesons in Photoproduction and Deep Inelastic Scattering at HERA},''
  \href{http://dx.doi.org/10.1140/epjc/s10052-010-1376-5}{{\em Eur. Phys. J. C}
  {\bfseries 68} (2010) 401--420},
  \href{http://arxiv.org/abs/1002.0234}{{\ttfamily arXiv:1002.0234 [hep-ex]}}.

\bibitem{Abramowicz:2012dh}
{\bfseries ZEUS} Collaboration, H.~Abramowicz {\em et~al.}, ``{Measurement of
  inelastic $J/\psi$ and $\psi^\prime$ photoproduction at HERA},''
  \href{http://dx.doi.org/10.1007/JHEP02(2013)071}{{\em JHEP} {\bfseries 02}
  (2013) 071},
\href{http://arxiv.org/abs/1211.6946}{{\ttfamily arXiv:1211.6946 [hep-ex]}}.

\bibitem{Jung:1992uj}
H.~Jung, G.~A. Schuler, and J.~Terron, ``{J / psi production mechanisms and
  determination of the gluon density at HERA},''
\href{http://dx.doi.org/10.1142/S0217751X92003604}{{\em Int. J. Mod. Phys.}
  {\bfseries A7} (1992) 7955--7988}.

\bibitem{Bacchetta:2018ivt}
A.~Bacchetta, D.~Boer, C.~Pisano, and P.~Taels, ``{Gluon TMDs and NRQCD matrix
  elements in $J/\psi$ production at an EIC},''
  \href{http://dx.doi.org/10.1140/epjc/s10052-020-7620-8}{{\em Eur. Phys. J.}
  {\bfseries C80} no.~1, (2020) 72},
\href{http://arxiv.org/abs/1809.02056}{{\ttfamily arXiv:1809.02056 [hep-ph]}}.

\bibitem{DAlesio:2019qpk}
U.~D'Alesio, F.~Murgia, C.~Pisano, and P.~Taels, ``{Azimuthal asymmetries in
  semi-inclusive $J/\psi\,+\,\mathrm{jet}$ production at an EIC},''
  \href{http://dx.doi.org/10.1103/PhysRevD.100.094016}{{\em Phys. Rev.}
  {\bfseries D100} no.~9, (2019) 094016},
\href{http://arxiv.org/abs/1908.00446}{{\ttfamily arXiv:1908.00446 [hep-ph]}}.

\bibitem{Kishore:2019fzb}
R.~Kishore, A.~Mukherjee, and S.~Rajesh, ``{Sivers asymmetry in the
  photoproduction of a $J/\psi$ and a jet at the EIC},''
  \href{http://dx.doi.org/10.1103/PhysRevD.101.054003}{{\em Phys. Rev.}
  {\bfseries D101} no.~5, (2020) 054003},
\href{http://arxiv.org/abs/1908.03698}{{\ttfamily arXiv:1908.03698 [hep-ph]}}.

\bibitem{Boer:2020bbd}
D.~Boer, U.~D'Alesio, F.~Murgia, C.~Pisano, and P.~Taels, ``{$J/\psi$ meson
  production in SIDIS: matching high and low transverse momentum},''
  \href{http://dx.doi.org/10.1007/JHEP09(2020)040}{{\em JHEP} {\bfseries 09}
  (2020) 040},
\href{http://arxiv.org/abs/2004.06740}{{\ttfamily arXiv:2004.06740 [hep-ph]}}.

\bibitem{Accardi:2012qut}
A.~Accardi {\em et~al.}, ``{Electron Ion Collider: The Next QCD Frontier}:
  {Understanding the glue that binds us all},''
  \href{http://dx.doi.org/10.1140/epja/i2016-16268-9}{{\em Eur. Phys. J. A}
  {\bfseries 52} no.~9, (2016) 268},
  \href{http://arxiv.org/abs/1212.1701}{{\ttfamily arXiv:1212.1701 [nucl-ex]}}.

\bibitem{Berger:1980ni}
E.~L. Berger and D.~L. Jones, ``{Inelastic Photoproduction of J/psi and Upsilon
  by Gluons},''
\href{http://dx.doi.org/10.1103/PhysRevD.23.1521}{{\em Phys. Rev.} {\bfseries
  D23} (1981) 1521--1530}.

\bibitem{Bodwin:1994jh}
G.~T. Bodwin, E.~Braaten, and G.~P. Lepage, ``{Rigorous QCD analysis of
  inclusive annihilation and production of heavy quarkonium},''
  \href{http://dx.doi.org/10.1103/PhysRevD.55.5853,
  10.1103/PhysRevD.51.1125}{{\em Phys. Rev.} {\bfseries D51} (1995)
  1125--1171}, \href{http://arxiv.org/abs/hep-ph/9407339}{{\ttfamily
  arXiv:hep-ph/9407339 [hep-ph]}}.
[Erratum: Phys. Rev.D55,5853(1997)].

\bibitem{Kramer:1995nb}
M.~Kraemer, ``{QCD corrections to inelastic J / psi photoproduction},''
  \href{http://dx.doi.org/10.1016/0550-3213(95)00568-4}{{\em Nucl. Phys.}
  {\bfseries B459} (1996) 3--50},
\href{http://arxiv.org/abs/hep-ph/9508409}{{\ttfamily arXiv:hep-ph/9508409
  [hep-ph]}}.

\bibitem{Artoisenet:2008fc}
P.~Artoisenet, J.~M. Campbell, J.~Lansberg, F.~Maltoni, and F.~Tramontano,
  ``{$\Upsilon$ Production at Fermilab Tevatron and LHC Energies},''
  \href{http://dx.doi.org/10.1103/PhysRevLett.101.152001}{{\em Phys. Rev.
  Lett.} {\bfseries 101} (2008) 152001},
  \href{http://arxiv.org/abs/0806.3282}{{\ttfamily arXiv:0806.3282 [hep-ph]}}.

\bibitem{Lansberg:2008gk}
J.~Lansberg, ``{On the mechanisms of heavy-quarkonium hadroproduction},''
  \href{http://dx.doi.org/10.1140/epjc/s10052-008-0826-9}{{\em Eur. Phys. J. C}
  {\bfseries 61} (2009) 693--703},
  \href{http://arxiv.org/abs/0811.4005}{{\ttfamily arXiv:0811.4005 [hep-ph]}}.

\bibitem{Lansberg:2009db}
J.~Lansberg, ``{Real next-to-next-to-leading-order QCD corrections to $J/\psi$
  and Upsilon hadroproduction in association with a photon},''
  \href{http://dx.doi.org/10.1016/j.physletb.2009.07.067}{{\em Phys. Lett. B}
  {\bfseries 679} (2009) 340--346},
  \href{http://arxiv.org/abs/0901.4777}{{\ttfamily arXiv:0901.4777 [hep-ph]}}.

\bibitem{Gong:2012ah}
B.~Gong, J.-P. Lansberg, C.~Lorce, and J.~Wang, ``{Next-to-leading-order QCD
  corrections to the yields and polarisations of J/Psi and Upsilon directly
  produced in association with a Z boson at the LHC},''
  \href{http://dx.doi.org/10.1007/JHEP03(2013)115}{{\em JHEP} {\bfseries 03}
  (2013) 115},
\href{http://arxiv.org/abs/1210.2430}{{\ttfamily arXiv:1210.2430 [hep-ph]}}.

\bibitem{Lansberg:2013qka}
J.-P. Lansberg and H.-S. Shao, ``{Production of $J/\psi + \eta_{c}$ versus
  $J/\psi + J/\psi$ at the LHC: Importance of Real $\alpha^{5}_{s}$
  Corrections},'' \href{http://dx.doi.org/10.1103/PhysRevLett.111.122001}{{\em
  Phys. Rev. Lett.} {\bfseries 111} (2013) 122001},
\href{http://arxiv.org/abs/1308.0474}{{\ttfamily arXiv:1308.0474 [hep-ph]}}.

\bibitem{Lansberg:2014swa}
J.-P. Lansberg and H.-S. Shao, ``{J/psi -pair production at large momenta:
  Indications for double parton scatterings and large $\alpha_s^5$
  contributions},''
  \href{http://dx.doi.org/10.1016/j.physletb.2015.10.083}{{\em Phys. Lett.}
  {\bfseries B751} (2015) 479--486},
\href{http://arxiv.org/abs/1410.8822}{{\ttfamily arXiv:1410.8822 [hep-ph]}}.

\bibitem{Artoisenet:2009xh}
P.~Artoisenet, J.~M. Campbell, F.~Maltoni, and F.~Tramontano, ``{$J/\psi$
  production at HERA},''
  \href{http://dx.doi.org/10.1103/PhysRevLett.102.142001}{{\em Phys. Rev.
  Lett.} {\bfseries 102} (2009) 142001},
  \href{http://arxiv.org/abs/0901.4352}{{\ttfamily arXiv:0901.4352 [hep-ph]}}.

\bibitem{Butenschoen:2009zy}
M.~Butenschoen and B.~A. Kniehl, ``{Complete next-to-leading-order corrections
  to J/psi photoproduction in nonrelativistic quantum chromodynamics},''
  \href{http://dx.doi.org/10.1103/PhysRevLett.104.072001}{{\em Phys. Rev.
  Lett.} {\bfseries 104} (2010) 072001},
\href{http://arxiv.org/abs/0909.2798}{{\ttfamily arXiv:0909.2798 [hep-ph]}}.

\bibitem{Chang:2009uj}
C.-H. Chang, R.~Li, and J.-X. Wang, ``{J/psi polarization in photo-production
  up-to the next-to-leading order of QCD},''
  \href{http://dx.doi.org/10.1103/PhysRevD.80.034020}{{\em Phys. Rev.}
  {\bfseries D80} (2009) 034020},
\href{http://arxiv.org/abs/0901.4749}{{\ttfamily arXiv:0901.4749 [hep-ph]}}.

\bibitem{Godbole:1995ie}
R.~M. Godbole, D.~P. Roy, and K.~Sridhar, ``{$J/\psi$ production via
  fragmentation at HERA},''
  \href{http://dx.doi.org/10.1016/0370-2693(96)00144-X}{{\em Phys. Lett.}
  {\bfseries B373} (1996) 328--333},
\href{http://arxiv.org/abs/hep-ph/9511433}{{\ttfamily arXiv:hep-ph/9511433
  [hep-ph]}}.

\bibitem{Kniehl:1997gh}
B.~A. Kniehl and G.~Kramer, ``{Charmonium production via fragmentation at DESY
  HERA},'' \href{http://dx.doi.org/10.1103/PhysRevD.56.5820}{{\em Phys. Rev.}
  {\bfseries D56} (1997) 5820--5833},
\href{http://arxiv.org/abs/hep-ph/9706369}{{\ttfamily arXiv:hep-ph/9706369
  [hep-ph]}}.

\bibitem{Berger:1982fh}
E.~L. Berger and D.~L. Jones, ``{Heavy Quark Contributions to Inelastic
  Photoproduction of the $J/\psi$, $\Upsilon$, and Other States},''
\href{http://dx.doi.org/10.1016/0370-2693(83)90203-4}{{\em Phys. Lett.}
  {\bfseries 121B} (1983) 61--64}.

\bibitem{Lansberg:2013wva}
J.~P. Lansberg and C.~Lorce, ``{Reassessing the importance of the
  colour-singlet contributions to direct $J/\psi + W$ production at the LHC and
  the Tevatron},'' \href{http://dx.doi.org/10.1016/j.physletb.2013.07.059,
  10.1016/j.physletb.2014.10.015}{{\em Phys. Lett.} {\bfseries B726} (2013)
  218--222}, \href{http://arxiv.org/abs/1303.5327}{{\ttfamily arXiv:1303.5327
  [hep-ph]}}.
[Erratum: Phys. Lett.B738,529(2014)].

\bibitem{Brodsky:1980pb}
S.~J. Brodsky, P.~Hoyer, C.~Peterson, and N.~Sakai, ``{The Intrinsic Charm of
  the Proton},''
\href{http://dx.doi.org/10.1016/0370-2693(80)90364-0}{{\em Phys. Lett.}
  {\bfseries 93B} (1980) 451--455}.

\bibitem{Qiu:2020xum}
J.-W. Qiu, X.-P. Wang, and H.~Xing, ``{Exploring $J/\psi$ production mechanism
  at the future Electron-Ion Collider},''
  \href{http://arxiv.org/abs/2005.10832}{{\ttfamily arXiv:2005.10832
  [hep-ph]}}.

\bibitem{Chang:1979nn}
C.-H. Chang, ``{Hadronic Production of $J/\psi$ Associated With a Gluon},''
\href{http://dx.doi.org/10.1016/0550-3213(80)90175-3}{{\em Nucl. Phys.}
  {\bfseries B172} (1980) 425--434}.

\bibitem{Baier:1983va}
R.~Baier and R.~Ruckl, ``{Hadronic Collisions: A Quarkonium Factory},''
\href{http://dx.doi.org/10.1007/BF01572254}{{\em Z. Phys.} {\bfseries C19}
  (1983) 251}.

\bibitem{Shao:2012iz}
H.-S. Shao, ``{HELAC-Onia: An automatic matrix element generator for heavy
  quarkonium physics},''
  \href{http://dx.doi.org/10.1016/j.cpc.2013.05.023}{{\em Comput. Phys.
  Commun.} {\bfseries 184} (2013) 2562--2570},
  \href{http://arxiv.org/abs/1212.5293}{{\ttfamily arXiv:1212.5293 [hep-ph]}}.

\bibitem{Shao:2015vga}
H.-S. Shao, ``{HELAC-Onia 2.0: an upgraded matrix-element and event generator
  for heavy quarkonium physics},''
  \href{http://dx.doi.org/10.1016/j.cpc.2015.09.011}{{\em Comput. Phys.
  Commun.} {\bfseries 198} (2016) 238--259},
  \href{http://arxiv.org/abs/1507.03435}{{\ttfamily arXiv:1507.03435
  [hep-ph]}}.

\bibitem{Kniehl:1996we}
B.~A. Kniehl, G.~Kramer, and M.~Spira, ``{Large p(T) photoproduction of
  $D^{*\pm}$ mesons in e p collisions},''
  \href{http://dx.doi.org/10.1007/s002880050591}{{\em Z. Phys. C} {\bfseries
  76} (1997) 689--700}, \href{http://arxiv.org/abs/hep-ph/9610267}{{\ttfamily
  arXiv:hep-ph/9610267}}.

\bibitem{Butenschoen:2011yh}
M.~Butenschoen and B.~A. Kniehl, ``{World data of $J/\psi$ production
  consolidate NRQCD factorization at NLO},''
  \href{http://dx.doi.org/10.1103/PhysRevD.84.051501}{{\em Phys. Rev. D}
  {\bfseries 84} (2011) 051501},
  \href{http://arxiv.org/abs/1105.0820}{{\ttfamily arXiv:1105.0820 [hep-ph]}}.

\bibitem{Pumplin:2002vw}
J.~Pumplin, D.~Stump, J.~Huston, H.~Lai, P.~M. Nadolsky, and W.~Tung, ``{New
  generation of parton distributions with uncertainties from global QCD
  analysis},'' \href{http://dx.doi.org/10.1088/1126-6708/2002/07/012}{{\em
  JHEP} {\bfseries 07} (2002) 012},
  \href{http://arxiv.org/abs/hep-ph/0201195}{{\ttfamily arXiv:hep-ph/0201195}}.

\bibitem{Aivazis:1993pi}
M.~Aivazis, J.~C. Collins, F.~I. Olness, and W.-K. Tung, ``{Leptoproduction of
  heavy quarks. 2. A Unified QCD formulation of charged and neutral current
  processes from fixed target to collider energies},''
  \href{http://dx.doi.org/10.1103/PhysRevD.50.3102}{{\em Phys. Rev. D}
  {\bfseries 50} (1994) 3102--3118},
  \href{http://arxiv.org/abs/hep-ph/9312319}{{\ttfamily arXiv:hep-ph/9312319}}.

\bibitem{Shao:2020kgj}
H.-S. Shao, ``{$J/\psi$ meson production in association with an open charm
  hadron at the LHC: A reappraisal},''
  \href{http://dx.doi.org/10.1103/PhysRevD.102.034023}{{\em Phys. Rev. D}
  {\bfseries 102} (2020) 034023},
  \href{http://arxiv.org/abs/2005.12967}{{\ttfamily arXiv:2005.12967
  [hep-ph]}}.

\bibitem{Maltoni:2012pa}
F.~Maltoni, G.~Ridolfi, and M.~Ubiali, ``{b-initiated processes at the LHC: a
  reappraisal},'' \href{http://dx.doi.org/10.1007/JHEP04(2013)095,
  10.1007/JHEP07(2012)022}{{\em JHEP} {\bfseries 07} (2012) 022},
  \href{http://arxiv.org/abs/1203.6393}{{\ttfamily arXiv:1203.6393 [hep-ph]}}.
[Erratum: JHEP04,095(2013)].

\bibitem{Buza:1996wv}
M.~Buza, Y.~Matiounine, J.~Smith, and W.~L. van Neerven, ``{Charm
  electroproduction viewed in the variable flavor number scheme versus fixed
  order perturbation theory},''
  \href{http://dx.doi.org/10.1007/BF01245820}{{\em Eur. Phys. J.} {\bfseries
  C1} (1998) 301--320},
\href{http://arxiv.org/abs/hep-ph/9612398}{{\ttfamily arXiv:hep-ph/9612398
  [hep-ph]}}.

\bibitem{Olness:1997yc}
F.~I. Olness, R.~J. Scalise, and W.-K. Tung, ``{Heavy quark hadroproduction in
  perturbative QCD},'' \href{http://dx.doi.org/10.1103/PhysRevD.59.014506}{{\em
  Phys. Rev.} {\bfseries D59} (1999) 014506},
\href{http://arxiv.org/abs/hep-ph/9712494}{{\ttfamily arXiv:hep-ph/9712494
  [hep-ph]}}.

\bibitem{Thorne:1997ga}
R.~S. Thorne and R.~G. Roberts, ``{An Ordered analysis of heavy flavor
  production in deep inelastic scattering},''
  \href{http://dx.doi.org/10.1103/PhysRevD.57.6871}{{\em Phys. Rev.} {\bfseries
  D57} (1998) 6871--6898},
\href{http://arxiv.org/abs/hep-ph/9709442}{{\ttfamily arXiv:hep-ph/9709442
  [hep-ph]}}.

\bibitem{Cacciari:1998it}
M.~Cacciari, M.~Greco, and P.~Nason, ``{The P(T) spectrum in heavy flavor
  hadroproduction},''
  \href{http://dx.doi.org/10.1088/1126-6708/1998/05/007}{{\em JHEP} {\bfseries
  05} (1998) 007},
\href{http://arxiv.org/abs/hep-ph/9803400}{{\ttfamily arXiv:hep-ph/9803400
  [hep-ph]}}.

\bibitem{Kramer:2000hn}
M.~Kraemer, F.~I. Olness, and D.~E. Soper, ``{Treatment of heavy quarks in
  deeply inelastic scattering},''
  \href{http://dx.doi.org/10.1103/PhysRevD.62.096007}{{\em Phys. Rev.}
  {\bfseries D62} (2000) 096007},
\href{http://arxiv.org/abs/hep-ph/0003035}{{\ttfamily arXiv:hep-ph/0003035
  [hep-ph]}}.

\bibitem{Tung:2001mv}
W.-K. Tung, S.~Kretzer, and C.~Schmidt, ``{Open heavy flavor production in QCD:
  Conceptual framework and implementation issues},''
  \href{http://dx.doi.org/10.1088/0954-3899/28/5/321}{{\em J. Phys.} {\bfseries
  G28} (2002) 983--996},
\href{http://arxiv.org/abs/hep-ph/0110247}{{\ttfamily arXiv:hep-ph/0110247
  [hep-ph]}}.

\bibitem{Thorne:2006qt}
R.~S. Thorne, ``{A Variable-flavor number scheme for NNLO},''
  \href{http://dx.doi.org/10.1103/PhysRevD.73.054019}{{\em Phys. Rev.}
  {\bfseries D73} (2006) 054019},
\href{http://arxiv.org/abs/hep-ph/0601245}{{\ttfamily arXiv:hep-ph/0601245
  [hep-ph]}}.

\bibitem{Forte:2010ta}
S.~Forte, E.~Laenen, P.~Nason, and J.~Rojo, ``{Heavy quarks in deep-inelastic
  scattering},'' \href{http://dx.doi.org/10.1016/j.nuclphysb.2010.03.014}{{\em
  Nucl. Phys.} {\bfseries B834} (2010) 116--162},
\href{http://arxiv.org/abs/1001.2312}{{\ttfamily arXiv:1001.2312 [hep-ph]}}.

\bibitem{Forte:2015hba}
S.~Forte, D.~Napoletano, and M.~Ubiali, ``{Higgs production in bottom-quark
  fusion in a matched scheme},''
  \href{http://dx.doi.org/10.1016/j.physletb.2015.10.051}{{\em Phys. Lett.}
  {\bfseries B751} (2015) 331--337},
\href{http://arxiv.org/abs/1508.01529}{{\ttfamily arXiv:1508.01529 [hep-ph]}}.

\bibitem{Forte:2016sja}
S.~Forte, D.~Napoletano, and M.~Ubiali, ``{Higgs production in bottom-quark
  fusion: matching beyond leading order},''
  \href{http://dx.doi.org/10.1016/j.physletb.2016.10.040}{{\em Phys. Lett.}
  {\bfseries B763} (2016) 190--196},
\href{http://arxiv.org/abs/1607.00389}{{\ttfamily arXiv:1607.00389 [hep-ph]}}.

\bibitem{Krauss:2017wmx}
F.~Krauss and D.~Napoletano, ``{Towards a fully massive five-flavor scheme},''
  \href{http://dx.doi.org/10.1103/PhysRevD.98.096002}{{\em Phys. Rev.}
  {\bfseries D98} no.~9, (2018) 096002},
\href{http://arxiv.org/abs/1712.06832}{{\ttfamily arXiv:1712.06832 [hep-ph]}}.

\bibitem{Forte:2018ovl}
S.~Forte, D.~Napoletano, and M.~Ubiali, ``{$Z$ boson production in bottom-quark
  fusion: a study of $b$-mass effects beyond leading order},''
  \href{http://dx.doi.org/10.1140/epjc/s10052-018-6414-8}{{\em Eur. Phys. J.}
  {\bfseries C78} no.~11, (2018) 932},
\href{http://arxiv.org/abs/1803.10248}{{\ttfamily arXiv:1803.10248 [hep-ph]}}.

\bibitem{Duhr:2020kzd}
C.~Duhr, F.~Dulat, V.~Hirschi, and B.~Mistlberger, ``{Higgs production in
  bottom quark fusion: Matching the 4- and 5-flavour schemes to third order in
  the strong coupling},''
\href{http://arxiv.org/abs/2004.04752}{{\ttfamily arXiv:2004.04752 [hep-ph]}}.

\bibitem{Sjostrand:2014zea}
T.~Sj\"ostrand, S.~Ask, J.~R. Christiansen, R.~Corke, N.~Desai, P.~Ilten,
  S.~Mrenna, S.~Prestel, C.~O. Rasmussen, and P.~Z. Skands, ``{An introduction
  to PYTHIA 8.2},'' \href{http://dx.doi.org/10.1016/j.cpc.2015.01.024}{{\em
  Comput. Phys. Commun.} {\bfseries 191} (2015) 159--177},
  \href{http://arxiv.org/abs/1410.3012}{{\ttfamily arXiv:1410.3012 [hep-ph]}}.

\bibitem{Aaron:2012cj}
{\bfseries H1} Collaboration, F.~Aaron {\em et~al.}, ``{Measurement of Beauty
  Photoproduction near Threshold using Di-electron Events with the H1 Detector
  at HERA},'' \href{http://dx.doi.org/10.1140/epjc/s10052-012-2148-1}{{\em Eur.
  Phys. J. C} {\bfseries 72} (2012) 2148},
  \href{http://arxiv.org/abs/1206.4346}{{\ttfamily arXiv:1206.4346 [hep-ex]}}.

\bibitem{Dulat:2015mca}
S.~Dulat, T.-J. Hou, J.~Gao, M.~Guzzi, J.~Huston, P.~Nadolsky, J.~Pumplin,
  C.~Schmidt, D.~Stump, and C.~Yuan, ``{New parton distribution functions from
  a global analysis of quantum chromodynamics},''
  \href{http://dx.doi.org/10.1103/PhysRevD.93.033006}{{\em Phys. Rev. D}
  {\bfseries 93} no.~3, (2016) 033006},
  \href{http://arxiv.org/abs/1506.07443}{{\ttfamily arXiv:1506.07443
  [hep-ph]}}.

\bibitem{Buckley:2014ana}
A.~Buckley, J.~Ferrando, S.~Lloyd, K.~Nordström, B.~Page, M.~Rüfenacht,
  M.~Schönherr, and G.~Watt, ``{LHAPDF6: parton density access in the LHC
  precision era},''
  \href{http://dx.doi.org/10.1140/epjc/s10052-015-3318-8}{{\em Eur. Phys. J. C}
  {\bfseries 75} (2015) 132}, \href{http://arxiv.org/abs/1412.7420}{{\ttfamily
  arXiv:1412.7420 [hep-ph]}}.

\bibitem{Ma:2008gq}
Y.-Q. Ma, Y.-J. Zhang, and K.-T. Chao, ``{QCD correction to $e^+e^- \to
  J/\psi+gg$ at B Factories},''
  \href{http://dx.doi.org/10.1103/PhysRevLett.102.162002}{{\em Phys. Rev.
  Lett.} {\bfseries 102} (2009) 162002},
\href{http://arxiv.org/abs/0812.5106}{{\ttfamily arXiv:0812.5106 [hep-ph]}}.

\bibitem{Brodsky:2009cf}
S.~J. Brodsky and J.-P. Lansberg, ``{Heavy-Quarkonium Production in High Energy
  Proton-Proton Collisions at RHIC},''
  \href{http://dx.doi.org/10.1103/PhysRevD.81.051502}{{\em Phys. Rev. D}
  {\bfseries 81} (2010) 051502},
  \href{http://arxiv.org/abs/0908.0754}{{\ttfamily arXiv:0908.0754 [hep-ph]}}.

\bibitem{Abe:2002rb}
{\bfseries Belle} Collaboration, K.~Abe {\em et~al.}, ``{Observation of double
  c anti-c production in $e^+ e^-$ annihilation at sqrt{s} approximately
  10.6-GeV},'' \href{http://dx.doi.org/10.1103/PhysRevLett.89.142001}{{\em
  Phys. Rev. Lett.} {\bfseries 89} (2002) 142001},
  \href{http://arxiv.org/abs/hep-ex/0205104}{{\ttfamily arXiv:hep-ex/0205104}}.

\bibitem{Pakhlov:2009nj}
{\bfseries Belle} Collaboration, P.~Pakhlov {\em et~al.}, ``{Measurement of the
  $e^+ e^- \to J/\psi\,c \bar c$ cross section at sqrt{s} \textasciitilde
  10.6-GeV},'' \href{http://dx.doi.org/10.1103/PhysRevD.79.071101}{{\em Phys.
  Rev. D} {\bfseries 79} (2009) 071101},
  \href{http://arxiv.org/abs/0901.2775}{{\ttfamily arXiv:0901.2775 [hep-ex]}}.

\bibitem{Zhang:2006ay}
Y.-J. Zhang and K.-T. Chao, ``{Double charm production e+ e- ---> J / psi + c
  anti-c at B factories with next-to-leading order QCD correction},''
  \href{http://dx.doi.org/10.1103/PhysRevLett.98.092003}{{\em Phys. Rev. Lett.}
  {\bfseries 98} (2007) 092003},
\href{http://arxiv.org/abs/hep-ph/0611086}{{\ttfamily arXiv:hep-ph/0611086
  [hep-ph]}}.

\bibitem{Gong:2009ng}
B.~Gong and J.-X. Wang, ``{Next-to-leading-order QCD corrections to e+e- -->
  J/psi(cc) at the B factories},''
  \href{http://dx.doi.org/10.1103/PhysRevD.80.054015}{{\em Phys. Rev.}
  {\bfseries D80} (2009) 054015},
\href{http://arxiv.org/abs/0904.1103}{{\ttfamily arXiv:0904.1103 [hep-ph]}}.

\bibitem{Baranov:2006dh}
S.~Baranov, ``{Topics in associated J/psi + c + anti-c production at modern
  colliders},'' \href{http://dx.doi.org/10.1103/PhysRevD.73.074021}{{\em Phys.
  Rev. D} {\bfseries 73} (2006) 074021}.

\bibitem{Artoisenet:2007xi}
P.~Artoisenet, J.~Lansberg, and F.~Maltoni, ``{Hadroproduction of $J/\psi$ and
  $\Upsilon$ in association with a heavy-quark pair},''
  \href{http://dx.doi.org/10.1016/j.physletb.2007.04.031}{{\em Phys. Lett. B}
  {\bfseries 653} (2007) 60--66},
  \href{http://arxiv.org/abs/hep-ph/0703129}{{\ttfamily arXiv:hep-ph/0703129}}.

\bibitem{Lansberg:2010vq}
J.~P. Lansberg, ``{QCD corrections to J/psi polarisation in pp collisions at
  RHIC},'' \href{http://dx.doi.org/10.1016/j.physletb.2010.10.054}{{\em Phys.
  Lett.} {\bfseries B695} (2011) 149--156},
\href{http://arxiv.org/abs/1003.4319}{{\ttfamily arXiv:1003.4319 [hep-ph]}}.

\bibitem{Chudakov:2016otl}
E.~Chudakov, D.~Higinbotham, C.~Hyde, S.~Furletov, Y.~Furletova, D.~Nguyen,
  M.~Stratmann, M.~Strikman, C.~Weiss, and R.~Yoshida, ``{Probing nuclear
  gluons with heavy quarks at EIC},''
  \href{http://dx.doi.org/10.22323/1.265.0143}{{\em PoS} {\bfseries DIS2016}
  (2016) 143}, \href{http://arxiv.org/abs/1608.08686}{{\ttfamily
  arXiv:1608.08686 [hep-ph]}}.

\bibitem{Furletova:INT}
Y.~Furletova. Talk at {INT} {S}eattle, {N}ov. 2nd, 2018.
\newblock
  \url{http://www.int.washington.edu/talks/WorkShops/int_18_3/People/Furletova_Y/Furletova.pdf}.

\bibitem{Hou:2017khm}
T.-J. Hou, S.~Dulat, J.~Gao, M.~Guzzi, J.~Huston, P.~Nadolsky, C.~Schmidt,
  J.~Winter, K.~Xie, and C.~P. Yuan, ``{CT14 Intrinsic Charm Parton
  Distribution Functions from CTEQ-TEA Global Analysis},''
  \href{http://dx.doi.org/10.1007/JHEP02(2018)059}{{\em JHEP} {\bfseries 02}
  (2018) 059}, \href{http://arxiv.org/abs/1707.00657}{{\ttfamily
  arXiv:1707.00657 [hep-ph]}}.

\bibitem{Sufian:2020coz}
R.~S. Sufian, T.~Liu, A.~Alexandru, S.~J. Brodsky, G.~F. de~Téramond, H.~G.
  Dosch, T.~Draper, K.-F. Liu, and Y.-B. Yang, ``{Constraints on
  charm-anticharm asymmetry in the nucleon from lattice QCD},''
  \href{http://dx.doi.org/10.1016/j.physletb.2020.135633}{{\em Phys. Lett.}
  {\bfseries B808} (2020) 135633},
\href{http://arxiv.org/abs/2003.01078}{{\ttfamily arXiv:2003.01078 [hep-lat]}}.

\bibitem{Nayak:2007mb}
G.~C. Nayak, J.-W. Qiu, and G.~F. Sterman, ``{Color transfer in associated
  heavy-quarkonium production},''
  \href{http://dx.doi.org/10.1103/PhysRevLett.99.212001}{{\em Phys. Rev. Lett.}
  {\bfseries 99} (2007) 212001},
\href{http://arxiv.org/abs/0707.2973}{{\ttfamily arXiv:0707.2973 [hep-ph]}}.

\bibitem{AbelleiraFernandez:2012cc}
{\bfseries LHeC Study Group} Collaboration, J.~Abelleira~Fernandez {\em
  et~al.}, ``{A Large Hadron Electron Collider at CERN: Report on the Physics
  and Design Concepts for Machine and Detector},''
  \href{http://dx.doi.org/10.1088/0954-3899/39/7/075001}{{\em J. Phys. G}
  {\bfseries 39} (2012) 075001},
  \href{http://arxiv.org/abs/1206.2913}{{\ttfamily arXiv:1206.2913
  [physics.acc-ph]}}.

\bibitem{Agostini:2020fmq}
{\bfseries LHeC, FCC-he Study Group} Collaboration, P.~Agostini {\em et~al.},
  ``{The Large Hadron-Electron Collider at the HL-LHC},''
  \href{http://arxiv.org/abs/2007.14491}{{\ttfamily arXiv:2007.14491
  [hep-ex]}}.

\end{thebibliography}\endgroup

\end{document}